\documentclass[twocolumn,trackchanges, tighten]{aastex631}

\usepackage{amsmath}

\def \OIII {[O\,{\sc iii}]}

\def\FeII{Fe\,{\sc ii}}

\begin{document}

\title{The size of the continuum emission region and its scaling relations with active galactic nucleus luminosity and the broad-line region size}

\author[0000-0001-9957-6349]{Amit Kumar Mandal}
\affiliation{Department of Physics $\&$ Astronomy, Seoul National University\\
Seoul 08826, Republic of Korea, jhwoo@snu.ac.kr}
\affiliation{Center for Theoretical Physics, Polish Academy of Sciences, Al. Lotnik\'ow 32/46, 02-668 Warsaw, Poland}

\author[0000-0002-8055-5465]{Jong-Hak Woo}
\affiliation{Department of Physics $\&$ Astronomy, Seoul National University\\
Seoul 08826, Republic of Korea, jhwoo@snu.ac.kr}

\author[0000-0002-2052-6400]{Shu Wang}
\affiliation{Department of Physics $\&$ Astronomy, Seoul National University\\
Seoul 08826, Republic of Korea, jhwoo@snu.ac.kr}







\begin{abstract}

We present a continuum lag analysis for a sample of 37 relatively high-luminosity active galactic nuclei (AGNs) 
from the Seoul National University AGN Monitoring Project (SAMP), utilizing the light curve data in $B$ and $V$ bands from SAMP and in $g,r,i$ bands from the Zwicky Transient Facility. 
We find that the inter-band lags ($\tau$) increase with wavelength (i.e., $\tau \propto \lambda^{\sim 4/3}$) as prescribed by the standard disk model (SSD), suggesting consistency with the “lamp-post” reprocessing model.  
We report that the size of the continuum emitting region (CER) normalized at 2500 {\AA} ($R_{2500}$) is a factor of $\sim$5 (i.e, $0.69\pm0.04$ dex) larger than predicted by SSD. 
By combining our new measurements with the re-measurements of the literature sample, we report a correlation between $R_{2500}$ and AGN continuum luminosity as $R_{2500} \, \propto \, L_{5100}^{0.58\pm0.03}$, which suggests  that the observed continuum could be composed of both the disk emission and the diffuse emission from the broad line region (BLR). The size of CER shows a tight relation with the size of H$\beta$ BLR with a sublinear slope (i.e., $R_{\text{BLR}} \, \propto \, R_{\text{2500}}^{0.87\pm0.07}$) and a scatter of 0.29 dex. This empirical relation offers a promising method for estimating single-epoch black hole masses, once established over a large dynamic range of AGN luminosity.

\end{abstract}

\keywords{Reverberation mapping (2019) --- Black holes (162) --- Active galactic nuclei (16) --- Quasars (1319)}


\section{Introduction} \label{sec:intro}

Active Galactic Nuclei (AGNs) are characterized by exceptional luminosity, emanating from a concentrated region surrounding a supermassive black hole (SMBH).
Mass-accreting supermassive black holes can form a geometrically thin, optically thick accretion disk, which emits thermal radiation in the X-ray to optical wavelength range with a peak in the UV  \citep{1973A&A....24..337S}.
However, various studies revealed a more complex structure of the accretion disk, which is closely tied to the Eddington ratio ($\lambda_{\text{Edd}}=L/L_{\text{Edd}}$) \citep[e.g.,][]{1988ApJ...332..646A, 1995ApJ...452..710N, 2013LRR....16....1A, 2014ARA&A..52..529Y}.
A number of accretion disk models have been proposed, ranging from the standard Shakura-Sunyaev disk  \citep[SSD;][]{1973A&A....24..337S} model and the slim disk  \citep[$\lambda_{\text{Edd}}\gtrsim 1$;][]{1988ApJ...332..646A} model, to more intricated models, i.e., the advection-dominated accretion flow \citep[$\lambda_{\text{Edd}}\ll 1$;][]{1995ApJ...452..710N} and the relativistic accretion disk surrounding a Kerr black hole \citep{2005ApJS..157..335L}.

In the framework of the SSD model, the radiation from the accretion disk is assumed to be a composite of multi-temperature blackbodies as the disk displays a temperature gradient with the disk radius (R) as $T(R) \propto (M\dot{M})^{1/4}R^{-3/4}$. 
Consequently, the wavelength-dependent disk radius ($R_{\lambda}$) 
follows a power-law relation as $R_{\lambda} \, \propto \, \lambda^{\beta}$, with $\beta = 4/3$.

The bulk of the accretion energy is liberated in close proximity to the black hole, in the form of radiation and extensive outflows. Therefore, comprehending the inner accretion flow and structure of the different emitting regions within the accretion disk is crucial for constraining AGN feedback. Although these scales are typically too small for direct spatial resolution, reverberation mapping \citep[RM;][]{1982ApJ...255..419B, 1993PASP..105..247P} offers an effective method to probe the disk structure, where time resolution can be utilized instead of spatial resolution. This disk mapping uses wavelength-dependent flux variability, often assumed to follow the lamp-post model \citep{1991ApJ...371..541K, 2007MNRAS.380..669C}. According to this model, X-rays from a hot corona above the disk illuminate the accretion disk, resulting in reprocessed radiations in the UV/optical. Consequently, longer wavelength photons originate from more extended segments of the disk, leading to time delays between continuum emissions across various wavelengths. Thus, the temporal behavior of continuum emission can elucidate the structure of the accretion disk.

Assuming that the reprocessed optical emission exhibits a delayed response to changes in the continuum at shorter wavelengths, the wavelength-dependent lags can be expressed as \citep{2021iSci...24j2557C}
\begin{equation} \label{fit_eq}
\tau =  \frac{R_{\lambda_{0}}}{c} \left[\left(\frac{\lambda}{\lambda_{0}}\right)^{\beta}-1\right]
\end{equation}
where $\lambda_{0}$ denotes the reference band wavelength, for which a band with most robust data is selected,
$R_{\lambda_{0}}$ is the size of the continuum emission region (CER) normalized at $\lambda_{0}$, $\beta$ is assumed to be 4/3 according to SSD model or a free parameter.

In recent years AGN monitoring campaigns with high cadence and refined analysis techniques have been advanced using both ground and space telescopes, reporting continuum reverberation time delays over X-ray, UV-optical to NIR wavelengths for an increasing number of AGNs  \citep{2005ApJ...622..129S, 2014MNRAS.444.1469M, 2014ApJ...788...48S, 2017ApJ...836..186J, 2018ApJ...862..123M, 2019ApJ...880..126H, 2020ApJS..246...16Y, 2020ApJ...903..112D, 2022MNRAS.511.3005J, 2022ApJ...940...20G, 2023arXiv230902499S, 2024ApJ...973..152E}.

The key findings from these continuum reverberation mapping studies can be summarized as follows: First, UV-optical-infrared lags exhibit a wavelength-dependent increase, following $\tau \, \propto \, \lambda^{4/3}$, which is consistent with the expectation from the SSD model. However, the size of the CER at 2500{\AA} (i.e., $R_{2500}$) is a factor of several larger than the predicted value from the SSD model \citep[e.g.,][]{2022ApJ...940...20G, 2023ApJ...947...62K}. 
It is suggested that this discrepancy is due to contribution of the diffuse continuum emission (DCE) originating from the Broad Line Region (BLR) \citep{2001ApJ...553..695K, 2022MNRAS.509.2637N, 2022ApJ...940...20G}. However, a recent study on Fairall 9 by \citet{2024ApJ...973..152E} found that the observed inter-band time delays from Swift observations cannot be fully explained by the diffuse continuum emission from the BLR alone, assuming a zero-lag contribution from the standard disk model as proposed by \citet{2022MNRAS.509.2637N}. In contrast, \citet{2024arXiv241003597J} successfully reproduced the observed UV-optical time delays in NGC 5548 by incorporating a BLR dynamical model, which accounts for radiation pressure acting on dust, alongside a standard accretion disk model with an optional inner hot flow and the lamp-post model for disk irradiation.
Second, the lag-spectrum of a few AGNs shows an excess around the 3646 {\AA} Balmer jump \citep{2018ApJ...857...53C}, while this  feature is not observed in other AGNs \citep[e.g., Mrk 817][]{2021ApJ...922..151K}. 
Third, the correlation of the size of CER with AGN luminosity is reported based on several nearby AGNs \citep{2022MNRAS.509.2637N},
which is similar to the H$\beta$ BLR size--luminosity ($R_{\text{BLR}}$--$L$) relationship. For example, \citet{2022ApJ...940...20G} claimed that the CER size at 5100 {\AA}, $R_{5100}$ correlates with optical luminosity as $R_{5100} \, \propto \, L_{5100}^{0.48}$ using a much larger sample of 49 AGNs, where $L_{5100}$ denotes AGN continuum luminosity at 5100{\AA}. Forth, \citet{2023ApJ...948L..23W} showed that $R_{5100}$ correlates with $R_{\text{BLR}}$ based on 21 AGNs, for which both $R_{5100}$ and H$\beta$ BLR size measurements are available from RM studies. These scaling relations can offer a black hole mass estimator for a large number of AGNs.


The aforementioned results from the previous studies are very encouraging and providing new aspects of the structure of AGN accretion disk. However, the current sample of AGNs with the measured size of the CER from reverberation mapping is still limited in terms of the sample size and dynamical range. While the disk mapping is less difficult since the required time baseline is relatively short even for high-luminosity AGNs, a long-term monitoring over a 5-10 year period is required for the H$\beta$ lag measurements for luminous AGNs.


To refine and better define the $R_{5100}$--$L_{5100}$ (and thus $R_{2500}$--$L_{5100}$) and $R_{\text{BLR}}$--$R_{5100}$ (and thus $R_{\text{BLR}}$--$R_{2500}$) relationships, it becomes imperative to investigate potential sources of systematic uncertainty and selection bias.  To achieve this, a comprehensive study using a large sample of AGNs, covering a suitable range of AGN luminosity, is necessary.

\begin{table*}[]
\centering
\movetableright= -20mm
 \caption{Properties of the finally selected AGNs from SAMP and Literature samples}
 \label{tab:samp_sample}

\fontsize{8pt}{8pt}\selectfont
\begin{tabular}{llcclcr} \hline \hline

Number & Name & $z$ & log$L_{5100}$ & $R_{\text{BLR}}$ & $M_{\text{BH}}$ & Ref.    
\\ 
 & &  & ($erg \, s^{-1}$) & light-days & ($\times 10^{8}M_{\odot}$)  \\
(1) & (2) & (3) & (4) & (5)  & (6) & (7) 
\\ \hline

 &  &  & SAMP sample&  &  & 
 \\ 

1 & Mrk~1501  & 0.089 & $44.09\pm0.02$ & $11.7^{+7.6}_{-8.9}$ & $0.50^{+0.30}_{-0.40}$ & 1  \\
2 & PG~0026$+$129  & 0.142 & $44.99\pm0.01$ & $13.0^{+30.2}_{-18.5}$ & $0.30^{+0.80}_{-0.50}$ & 1    \\
3 & PG~0052$+$251  & 0.154 & $44.75\pm0.02$ & $63.9^{+12.0}_{-10.9}$ & $2.40^{+0.50}_{-0.50}$ & 1    \\
4 & J0101$+$422  & 0.190 & $44.82\pm0.01$ & $76.1^{+13.2}_{-11.8}$ & $3.50^{+0.70}_{-0.70}$ & 1    \\
5 & J0140$+$234  & 0.320 & $45.13\pm0.01$ & $113.6^{+9.5}_{-10.2}$ & $2.00^{+0.30}_{-0.30}$ & 1   \\
6 & PG~0947$+$396  & 0.206 & $44.58\pm0.01$ & $36.7^{+9.5}_{-11.0}$ & $1.70^{+0.50}_{-0.60}$ & 1    \\
7 & J1026$+$523  & 0.259 & $44.48\pm0.01$ & $34.1^{+4.2}_{-4.0}$ & $0.40^{+0.10}_{-0.10}$ & 1    \\
8 & J1120$+$423  & 0.226 & $44.53\pm0.01$ & $44.4^{+15.6}_{-14.9}$ & $1.80^{+0.70}_{-0.70}$ & 1    \\
9 & PG~1121$+$422  & 0.225 & $44.84\pm0.01$ & $115.8^{+24.2}_{-20.2}$ & $1.10^{+0.30}_{-0.20}$ & 1    \\
10 & PG~1202$+$281  & 0.165 & $44.44\pm0.01$ & $38.5^{+9.1}_{-8.5}$ & $1.20^{+0.30}_{-0.30}$ & 1    \\
11 & J1217$+$333  & 0.178 & $44.18\pm0.02$ & $26.5^{+21.2}_{-20.7}$ & $0.70^{+0.50}_{-0.50}$ & 1    \\
12 & VIII~Zw~218  & 0.127 & $44.46\pm0.01$ & $63.3^{+16.4}_{-15.4}$ & $1.60^{+0.40}_{-0.40}$ & 1    \\
13 & PG~1322$+$659  & 0.168 & $44.81\pm0.01$ & $49.3^{+19.2}_{-16.6}$ & $1.10^{+0.40}_{-0.40}$ & 1 \\
14 & J1415$+$483  & 0.275 & $44.64\pm0.01$ & $25.3^{+11.8}_{-11.1}$ & $0.80^{+0.40}_{-0.30}$ & 1 \\
15 & PG~1440$+$356  & 0.079 & $44.56\pm0.01$ & $50.6^{+16.9}_{-20.9}$ & $0.40^{+0.20}_{-0.20}$ & 1 \\
16 & J1456$+$380  & 0.283 & $44.73\pm0.01$ & $77.8^{+9.4}_{-8.7}$ & $4.30^{+0.70}_{-0.70}$ & 1 \\
17 & J1526$+$275  & 0.231 & $44.82\pm0.01$ & $63.9^{+10.3}_{-9.3}$ & $4.20^{+1.20}_{-1.80}$ & 1 \\
18 & J1540$+$355  & 0.164 & $44.45\pm0.01$ & $57.9^{+18.1}_{-14.6}$ & $0.60^{+0.20}_{-0.20}$ & 1 \\
19 & J1619$+$501  & 0.234 & $44.42\pm0.01$ & $32.3^{+6.8}_{-6.7}$ & $1.80^{+0.40}_{-0.40}$ & 1 \\
20 & PG~2349$-$014  & 0.174 & $44.66\pm0.02$ & $55.9^{+11.0}_{-10.8}$ & $6.30^{+1.40}_{-1.40}$ & 1 \\
 \hline \\
21 & PG~0003$+$158  & 0.451 & $45.73\pm0.01^a$ & ... & $9.11\pm3.01^b$ & ... \\
22 & PG~0043$+$039  & 0.385 & $45.36\pm0.01^a$ & ... & $4.53\pm1.49^b$ & ... \\
23 & J0848$+$282  & 0.198 & $44.15\pm0.01^a$ & ... & $1.53\pm0.50^b$ & ... \\
24 & J1059$+$665  & 0.340 & $44.90\pm0.01$ & ... & $2.53\pm0.83^b$ & 1 \\
25 & PG~1114$+$445  & 0.144 & $44.53\pm0.01^a$ & ... & $2.08\pm0.68^b$ & ... \\
26 & J1152$+$453  & 0.211 & $44.34\pm0.01^a$ & ... & $3.26\pm1.06^b$ &  ... \\
27 & J1203$+$455  & 0.343 & $44.99\pm0.01$ & ... & $4.45\pm1.46^b$ & 1  \\
28 & J1207$+$262  & 0.324 & $44.86\pm0.01^a$ & ... & $0.54\pm0.18^b$ & ... \\
29 & J1220$+$405  & 0.222 & $44.15\pm0.01^a$ & ... & $2.49\pm0.81^b$ & ... \\
30 & PG~1354$+$213  & 0.300 & $44.73\pm0.01^a$ & ... & $4.80\pm1.57^b$ & ... \\
31 & J1408$+$630  & 0.261 & $44.57\pm0.01^a$ & ... & $3.60\pm1.17^b$ & ... \\
32 & J1453$+$343  & 0.209 & $44.30\pm0.01^a$ & ... & $2.85\pm0.93^b$ & ... \\
33 & J1515$+$480  & 0.312 & $45.20\pm0.01^a$ & ... & $2.47\pm0.81^b$ & ... \\
34 & J1935$+$531  & 0.248 & $45.14\pm0.01$ & ... & $4.45\pm1.49^b$ & 1 \\
35 & PG~2130$+$099  & 0.063 & $44.38\pm0.01^a$ & ... & $0.45\pm0.15^b$ & ... \\
36 & PG~2214$+$139  & 0.066 & $44.44\pm0.01^a$ & ... & $2.02\pm0.66^b$ & ... \\
37 & PG~2251$+$113  & 0.326 & $45.49\pm0.01$ & ... & $5.11\pm1.69^b$ & 1 \\
 \hline  \hline
 &  &  & Literature sample&  &  & 
 \\ 

38 & Ark~120 & 0.033 & $43.81\pm0.25$ & $39.5^{+8.5}_{-7.8}$ & $1.50^{+0.19}_{-0.19}$ & 2,3 \\
39 & Fairall~9 & 0.046 & $43.92\pm0.05$ & $17.4^{+3.2}_{-4.3}$ & $2.55^{+0.56}_{-0.56}$ & 2,3 \\
40 & MCG~$+08-11-011$ & 0.021 & $43.28\pm0.05$ & $15.7^{+0.5}_{-0.5}$ & $0.28^{+0.31}_{-0.31}$ & 2,4  \\
41 & Mrk~110 & 0.035 & $43.60\pm0.02$ & $25.6^{+8.9}_{-7.2}$ & $0.25^{+0.06}_{-0.06}$ & 2,3  \\
42 & Mrk~142 & 0.045 & $43.53\pm0.04$ & $6.4^{+7.3}_{-3.4}$ & $0.02^{+0.01}_{-0.02}$ & 2,5,6  \\
43 & Mrk~509 & 0.035 & $44.13\pm0.03$ & $79.6^{+6.1}_{-5.4}$ & $1.43^{+0.12}_{-0.12}$ & 2,3  \\
44 & Mrk~817 & 0.031 & $43.68\pm 0.09$ & $19.9^{+9.9}_{-6.7}$ & $0.44^{+0.11}_{-0.11}$ & 2,5,7  \\

45 & NGC~2617 & 0.014 & $42.61\pm 0.10$ & $4.3^{+1.1}_{-1.3}$ & $0.32^{+0.35}_{-0.35}$ & 2,4  \\
46 & NGC~4151 & 0.003 & $42.31\pm 0.06$ & $6.8^{+0.5}_{-0.6}$ & $0.21^{+0.06}_{-0.06}$ & 2,8  \\
47 & NGC~4593 & 0.008 & $42.56\pm 0.37$ & $4.0^{+0.8}_{-0.7}$ & $0.08^{+0.02}_{-0.02}$ & 2,5,15 \\
48 & NGC~5548 & 0.017 & $43.24\pm 0.19$ & $13.9^{+11.2}_{-6.2}$ & $0.52^{+0.13}_{-0.13}$ & 2,16 \\
49 & Mrk~335 & 0.026 & $43.70\pm 0.08$ & $14.1^{+0.4}_{-0.4}$ & $0.25^{+0.03}_{-0.03}$ & 9 \\
50 & IRAS~09149$-$6206 & 0.057 & $44.53\pm 0.15$ & ... & $1.00^{+0.90}_{-0.90}$ & 10,11  \\
51 & Mrk~876 & 0.129 & $44.71\pm 0.03$ & $40.1^{+15.0}_{-15.2}$ & $2.79^{+1.29}_{-1.29}$ & 3,12 \\
52 & PG~1119+120 & 0.050 & $44.09\pm 0.05$ & $16.6^{+3.9}_{-4.5}$ & $0.09^{+0.09}_{-0.09}$ & 3,13  \\
53 & NGC~4395 & 0.001 & $39.76\pm 0.06$ & $0.058^{+0.010}_{-0.010}$ & $0.0002^{+0.0002}_{-0.0002}$ &2,14,17 \\
54 & PG~2130$+$099  & 0.063 & $44.38\pm0.01^a$ & ... & $0.45\pm0.15^b$ & ... \\

\hline

\end{tabular}

\vspace{0.05cm}
\begin{minipage}{\textwidth}
\footnotesize
\noindent \textbf{Note.} Columns are (1) serial number, (2) object Name,  (3) redshift, (4) AGN luminosities at 5100 {\AA}. The values marked by 'a' are derived in this study. (5) H$\beta$ BLR sizes in the rest-frame except for NGC 4395 of which H$\alpha$ BLR size is used, (6) BH masses from reverberation-mapping, while the single-epoch BH masses estimated in this study are marked with 'b', and (7) Ref. (references for luminosity, BLR sizes and BH mass): 1 \citet{2024ApJ...962...67W}, 2 \citet{2023ApJ...948L..23W}, 3 \citet{2004ApJ...613..682P}, 4 \citet{2017ApJ...840...97F}, 5 \citet{2019ApJ...886...42D}, 6 \citet{2018ApJ...869..137L}, 7 \citet{2010ApJ...721..715D}, 8 \citet{2018ApJ...866..133D}, 9 \citet{2012ApJ...755...60G}, 10 \citet{2023MNRAS.525.4524G}, 11 \citet{2020A&A...643A.154G}, 12 \citet{2013ApJ...767..149B}, 13 \citet{2023MNRAS.523..545D}, 14 \citet{2020ApJ...892...93C}, 15 \citet{2006ApJ...653..152D}, 16 \citet{2016ApJ...821...56F}, 17 \citet{2024ApJ...976..116P}.  
\end{minipage}
\end{table*}

In this study, we focus on a unique sample of type 1 AGNs, for which the size of H$\beta$ BLR is already available from the literature. For 37 AGNs, which is carefully selected from the Seoul National University AGN Monitoring Project (SAMP), we perform continuum reverberation mapping, and investigate the physical properties of the accretion disk. For the combined sample of AGNs with available H$\beta$ BLR size, we investigate the correlation of the CER size with optical luminosity and the size of BLR. 
The paper is structured as follows: Section \ref{sec:obs} describes the sample selection and the data. Section \ref{sec:analysis} presents the time series analysis and lag quality assessment. Results are presented in Section \ref{sec:result}, followed by a discussion and summary in Section \ref{sec:summ}.

\section{Sample and data}
\label{sec:obs}

\subsection{SAMP AGNs}

We selected a sample of 44 luminous AGNs from the Seoul National University AGN Monitoring Project \citep{2019JKAS...52..109W}, which was carried out to measure the size of the BLR and its scaling relation with AGN luminosity, focusing on high-luminosity AGNs. As described by \citet{2019JKAS...52..109W}, an initial sample of 100 AGNs with the monochromatic luminosity at 5100{\AA} $L_{5100} > 10^{44.0} \, erg \, s^{-1}$ out to $z \sim 0.4$ was selected. 
Among the initial sample of 100 AGNs, 29 targets were excluded based on the low variability and the intensity of the {\OIII} emission line by the end of 2016 while 71 AGNs were monitored for several years for photometry and spectroscopy.
By further selecting the best candidates with high variability amplitude and strong variability patterns in the light curve, 
a final sample of 32 AGNs were continuously monitored for six years from 2016 to 2021. The reverberation-mapping measurements of the H$\beta$ lag of 32 AGNs were reported by \citet{2024ApJ...962...67W}. 

Since only two bands ($B$ and $V$) are available from SAMP, we incorporated photometry data from the Zwicky Transient Facility survey \citep[ZTF;][]{2019PASP..131a8002B}, which commenced in 2018. 
Thus, we finally selected a sample of 44 AGNs from the SAMP, which were observed during 2018 - 2021 and have more than 100 epochs in the ZTF$-g$ band light curve during the period of 2018 to 2021. Note that the H$\beta$ BLR size was reported for 23 AGNs \citep{2024ApJ...962...67W}, while the H$\beta$ lag was not measured for the other 21 AGNs.


\begin{figure*}
\centering
\includegraphics[scale=0.69]{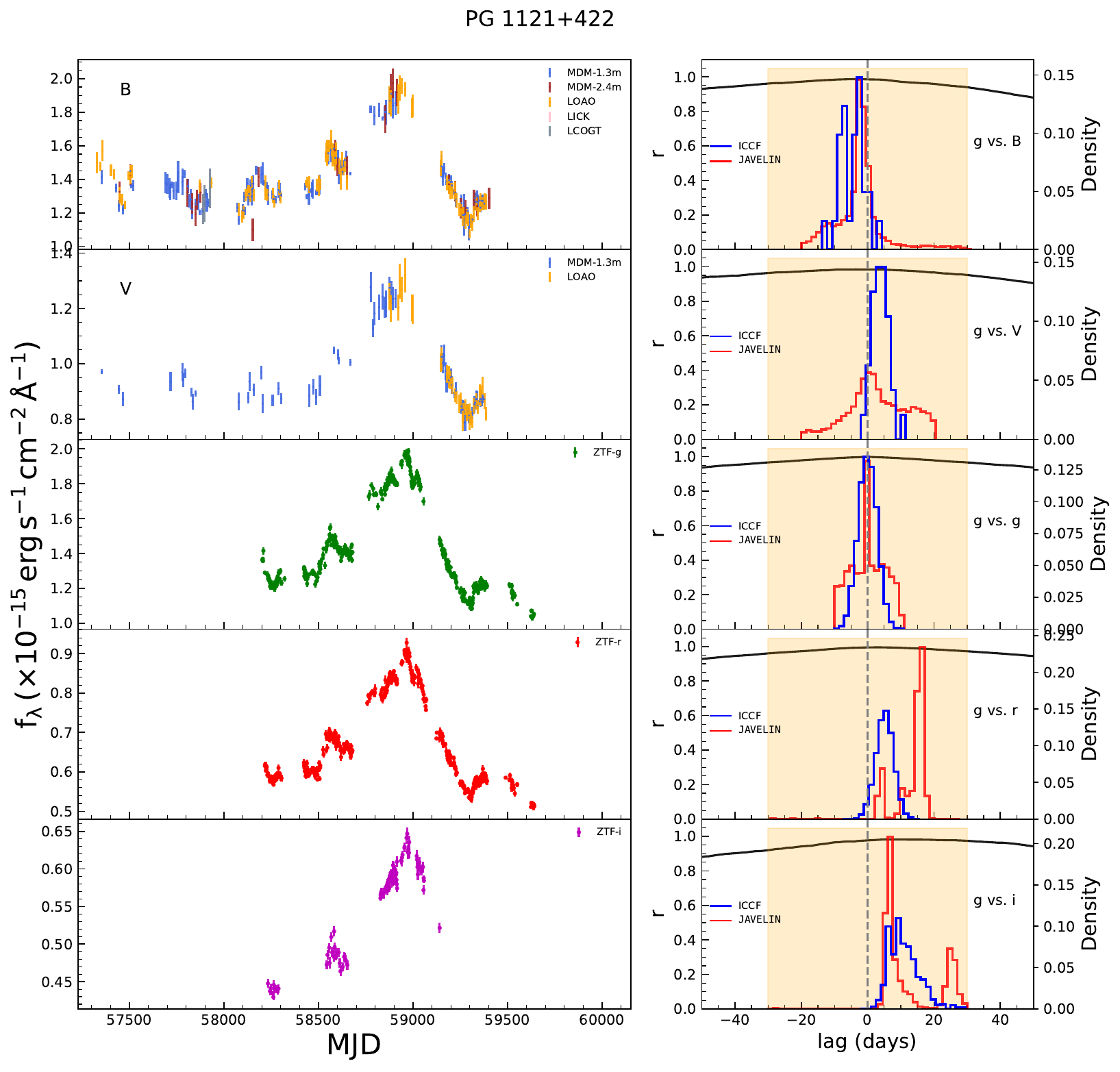}

\caption{Left: Example of the continuum light curves of the target PG 1121+422 in $B$, and $V-$ bands from the SAMP, and in $g,r,i-$bands from the ZTF survey. Right: Distribution of the cross correlation coefficient with respect to the ZTF$-g$ band (solid black line) using the ICCF. The cross-correlation centroid distribution (CCCD) obtained from ICCF is shown with the blue histogram, while the red histogram represents the lag probability distribution from {\tt JAVELIN}. The reference with $\tau$=0 day is denoted with a vertical dashed gray line. The orange shaded region presents the lag search window, which is set to $\pm$ 30 days. The full set of light curve plots for the SAMP targets is available online.}
\label{fig:light_ccf}
\end{figure*}

\begin{figure}
\centering
\includegraphics[scale=0.6]{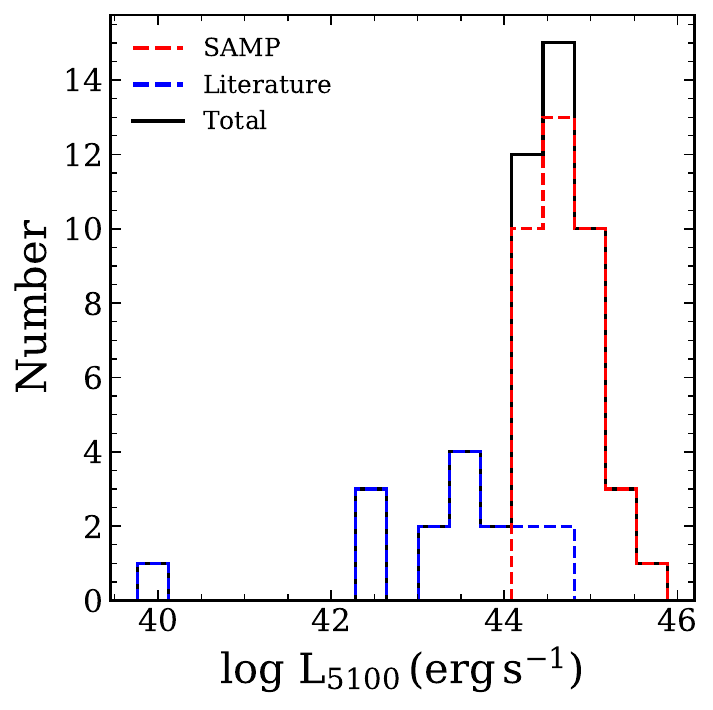}

\caption{Distribution of AGN luminosity at 5100 {\AA} for the SAMP sample (red line), the Literature sample (blue line), and the total sample (black line).}
\label{fig:dist}
\end{figure}

\subsection{SAMP Lightcurves}

The photometry data for the SAMP targets were obtained using multiple telescopes, including the MDM 1.3m and 2.4m telescopes located in Kitt Peak, Tucson, Arizona, USA, the Lemmonsan Optical Astronomy Observatory (LOAO) 1m telescope on Mt. Lemmon Optical Observatory, Tucson, Arizona, USA, the Lick Observatory 1m telescope located at Mt. Hamilton, California, USA, the Las Cumbres Observatory Global Telescope (LCOGT) network, and the Deokheung Optical Astronomy Observatory (DOAO) 1m telescope. The monitoring of continuum variability was performed with a typical cadence of 4-6 days using the $B$ and $V-$band filters \citep[for details see][]{2019JKAS...52..109W}.

A systematic offset between the light curves obtained from different telescopes was evident, presumably stemming from the fluctuations in weather conditions and variations in the filter properties. Thus, we inter-calibrated the obtained light curves using the python based software {\tt PyCALI} \footnote{\url{https://github.com/LiyrAstroph/PyCALI}}\citep{2014ApJ...786L...6L} as we performed in our previous works \citep[][in press]{2024ApJ...962...67W, 2024ApJS..275...13W}. {\tt PyCALI} employs Bayesian statistics to determine the best scaling factors for inter-calibration, by modeling each light curve using the damped random walk \citep[DRW;][]{2009ApJ...698..895K, 2010ApJ...721.1014M} model to describe AGN variability. We used light curve from the MDM 1.3m telescope as the reference to scale the other light curves since the MDM 1.3m light curve offered the most extensive dataset, with epochs that were evenly distributed over the monitoring time baseline. During the inter-calibration process, we added systematic uncertainties into the light curve of each telescope.

\subsection{ZTF Light Curves}

We collected the photometry data for the SAMP targets from the ZTF Data Release 14 (DR14). The ZTF light curves are available in the $g,r,i-$ bands with a cadence of approximately 2 to 4 days. 
As the ZTF employs point-source-function-fit (PSF) photometry to construct the light curves, we specifically selected the ZTF data that satisfied the condition {\bf catflags=0}, ensuring that we excluded data potentially contaminated by clouds, moonlight, or poor seeing conditions. We converted the ZTF magnitudes into fluxes by utilizing the zero-points provided by the Spanish Virtual Observatory (SVO) filter profile service \footnote{\url{http://svo2.cab.inta-csic.es/theory/fps/index.php?mode=browse&gname=Palomar&gname2=ZTF&asttype=}}. In cases where multiple observations were carried out during a single night, we computed an average flux.
An example of light curves is shown in Figure \ref{fig:light_ccf}.

\subsection{Literature sample}

In addition to the SAMP AGNs, we collected continuum lag measurements from the literature. We note that 17 AGNs were
intensively studied using multi-wavelength data from X-ray to optical wavelength and high quality lag measurements for these AGNs were previously reported \citep{2016ApJ...821...56F, 2017ApJ...840...41E, 2018ApJ...857...53C, 2018ApJ...854..107F, 2019ApJ...870..123E, 2020MNRAS.494.1165L, 2020MNRAS.498.5399H, 2020ApJ...896....1C, 2021ApJ...922..151K, 2021MNRAS.504.4337V, 2022ApJ...934L..37M, 2023ApJ...947...62K, 2023MNRAS.525.4524G, 2023MNRAS.523..545D, 2023ApJ...953..137M}. We refer
these AGNs as Literature sample. Note that we performed the disk-mapping analysis for Literature sample  as described in Section \ref{sec_disk}.

For a consistency check, we also used the sample of AGNs with continuum lag measurements from \citet{2022ApJ...940...20G}, who
utilized ZTF $g,r,i-$ band light curves to perform disk-mapping analysis. We collected the high-quality sub-sample of 94 AGNs, which were identified based on lag uncertainties, consistency between ICCF and {\tt JAVELIN} measurements, and lag reliability in their work. 
For a consistency, we further applied the same criteria, i.e., $r_{max} \geq 0.6$ and $p(r_{max})_{\tau>0} \leq 0.1$ as adopted for the SAMP sample (see Section \ref{sec:lag_qual}), finalizing a sample of 84 AGNs, which we define as Guo+22 sample. 
Note that since 3 band data are available for this sample, only 2 inter-band lag measurements, i.e., between $g$ and $r$ as well as $g$ and $i$ are obtained. As the fitting of the lag-spectrum with only two data points is relatively uncertain, we use Guo+22 sample only for a consistency check in the following analysis.

\subsection{AGN luminosity and mass}

We obtained black hole mass and AGN luminosity of the total sample to further investigate any dependency of the lag and scaling relations. 
For 23 SAMP AGNs, we adopted the AGN continuum luminosity at 5100 {\AA}, $L_{5100}$ from \citet{2024ApJ...962...67W}, which were measured based on spectral decomposition analysis to separate the AGN continuum from the stellar continuum. For the remaining AGNs in the SAMP sample, we utilized the spectra obtained from the SAMP and performed the same spectral decomposition to measure $L_{5100}$ as done by \citet{2024ApJ...962...67W}.
In the case of Literature sample we collected the AGN luminosity $L_{5100}$ from the previous reverberation studies, while for the Guo+22 sample, we adopted $L_{5100}$ from \citet{2022ApJ...940...20G}. 
In Figure \ref{fig:dist} we present the luminosity distribution of the SAMP sample and the Literature sample, which ranges over four orders of magnitude, with the SAMP AGNs covering the high luminosity rigime.

We also obtained reverberation-based black hole mass ($M_{\text{BH}}$) or single-epoch $M_{\text{BH}}$ for the total sample. 
First, we collected the $M_{\text{BH}}$ for 23 SAMP targets based on the H$\beta$ reverberation study by \citet{2024ApJ...962...67W}.
For the remaining 21 SAMP targets, we determined  single-epoch $M_{\text{BH}}$ with the FWHM of the H$\beta$ emission line and  $L_{5100}$ from the SAMP using the the H$\beta$ BLR size - luminosity relation reported by \citet{2024ApJ...962...67W} along with the virial factor of $f_{\text{BLR}} = 1.12$ \citep{2015ApJ...801...38W}. 
For Literature sample, we collected reverberation-based $M_{\text{BH}}$ from the literature.
We present the properties of the SAMP sample and the Literature sample in Table \ref{tab:samp_sample}.

\section{Analysis}
\label{sec:analysis}

\subsection{Continuum lag analysis of SAMP sample}

In this section, we present the lag analysis of the SAMP AGNs using the continuum light curves. All forty-four AGNs in the SAMP sample show significant flux variability with the fractional variability $F_{var}$ ranging from 0.07 to 0.21 in the $B-$band,
and present synchronized flux variations across all optical bands.

\subsubsection{Continuum lag measurements}

To determine the time lag between the light curves in different bands, we follow the method adopted in our previous study \citep[for details, see][]{2024ApJ...962...67W}. We employ the Interpolated Cross-Correlation Function \citep[ICCF;][]{1986ApJ...305..175G, 1987ApJS...65....1G}, using the PyCCF Python package \citep{1998PASP..110..660P, 2018ascl.soft05032S}.
This process includes linear interpolation of the two light curves and calculation of the cross-correlation coefficient (r). We set the lag search window to $\pm 30$ days, which is sufficiently large relative to the expected continuum lags  \citep[i.e., less than 10 days,][]{2022MNRAS.511.3005J} as it probes the region nearest to the thermal peak of the hot accretion disk \citep{2023ApJ...947...62K}.
To estimate the uncertainty of the lag, we adopt a model-independent Monte Carlo simulation based on flux randomization (FR) and random subset selection (RSS) as described in \citet{1998PASP..110..660P, 1999ApJ...526..579W}; and  \citet{2004ApJ...613..682P}. This simulation is repeated for 5000 iterations, and the centroid of the lag is determined in each iteration. From these results, a cross-correlation centroid distribution (CCCD) is constructed. The centroid lags are calculated from points above 80\% of the maximum of the cross-correlation function ($r_{max}$). The lag is determined from the median of the CCCD, with the lower and upper uncertainties in the measured lag corresponding to the 15.9 and 84.1 percentiles of the distribution. This is equivalent to a $1\sigma$ error in the case of a Gaussian distribution. We use the ZTF$-g$ band light curve as the reference light curve due to its superior sampling frequency and the highest S/N. We used the overlapped time baseline between the $g-$band and the other band light curves for determining the lags.

As a consistency check, we employ {\tt JAVELIN} \citep{2011ApJ...735...80Z} for the lag analysis to compare with the ICCF measurements.
The initial step of the {\tt JAVELIN} analysis is to model the reference light curve (i.e., ZTF$-g$) using the DRW model, which consists of two free parameters: amplitude ($\sigma_d$) and the time scale of variability ($\tau_d$). To retrieve the reprocessed continuum light curve, a top-hat transfer function is convolved with the reference light curve. We apply the same search window as used in the ICCF analysis to determine the lag. We adopt a Markov Chain Monte Carlo (MCMC) method to find the best-fitting model by maximizing the likelihood. The lag is finally derived from the median of the lag probability distribution around the most prominent peak in the distribution.

We find small inconsistency between the lags obtained with the ICCF and {\tt JAVELIN} methods, with an rms scatter of 0.47, 0.61, 0.62, and 0.63 dex for the lags between $g$ vs. $B$ ($\tau_{gB}$), $g$ vs. $V$ ($\tau_{gV}$), $g$ vs. $r$ ($\tau_{gr}$), and $g$ vs. $i$ ($\tau_{gi}$), respectively. This inconsistency is primarily attributed to the broad distributions of the lag posteriors and the sparsely sampled ZTF$-i$ band light curves with limited data points.  Further, we assess the consistency between lag measurements obtained from ICCF and {\tt JAVELIN} in Section \ref{sec:sim} (see Figure \ref{fig:sim_icfjav}). Our simulation indicates that while both methods yield consistent lag estimates within the signal-to-noise ratio (SNR) and cadence range of our light curves, {\tt JAVELIN} fails to recover the actual lags when the light curves contain larger seasonal gaps ($>$150 days). Therefore, to be consistent with previous studies of the H$\beta$ lag measurements of the SAMP AGNs, we adopt the results with the ICCF for the following analysis while we also present the results based on {\tt JAVELIN}.


\begin{table*}[]
\centering
 \movetableright= -65mm
 \caption{Time lag measurements of the SAMP sample}
 \label{tab:samp_laganalysis}

\resizebox{19cm}{!}{
\fontsize{20pt}{20pt}\selectfont
\begin{tabular}{lcccccccccccccr} \hline \hline

Number & Name & $\Delta t_{B}$ & $\Delta t_{V}$ & $\Delta t_{g}$ & $\Delta t_{r}$ & $\Delta t_{i}$ & $r_{max,gB}$  & $\tau_{gB,ICCF}$  & $r_{max,gV}$  & $\tau_{gV,ICCF}$ & $r_{max,gr}$  & $\tau_{gr,ICCF}$ & $r_{max,gi}$  & $\tau_{gi,ICCF}$ 
\\ 
 & & days & days & days & days & days & &  days & & days &  & days & & days\\
(1) & (2) & (3) & (4) & (5)  & (6) & (7) & (8) &  (9) & (10) & (11) & (12) & (13) & (14) & (15)
\\ \hline

1 & Mrk~1501 & 5.9 & 6.9 & 2.0 & 2.0 & 4.0 & 0.97  & $-4.0^{+4.5}_{-1.0}$  & 0.98 & $1.5^{+1.0}_{-3.5}$ & 0.97 & $3.9^{+1.5}_{-1.4}$  & 0.93 & $7.0^{+10.5}_{-4.5}$ \\
2 & PG~0026$+$129 & 4.9 & 5.9 & 2.0 & 2.0 & 4.0 & 0.92 & $1.8^{+10.9}_{-11.2}$  & 0.96 & $4.5^{+14.1}_{-13.6}$ & 0.97 & $14.1^{+2.8}_{-2.9}$ & 0.83 & $16.7^{+5.4}_{-6.3}$ \\
3 & PG~0052$+$251 & 5.9 & 6.1 & 2.1 & 2.0 & 4.5 & 0.99 & $-3.0^{+2.3}_{-3.0}$ & 0.98 & $3.9^{+5.4}_{-5.0}$ & 0.99 & $2.5^{+2.0}_{-1.0}$ & 0.90 & $18.2^{+4.2}_{-5.2}$ \\
4 & J0101$+$422 & 3.9 & 5.9 & 1.9 & 1.1 & 4.0 & 0.98 & $1.2^{+4.7}_{-4.9}$ & 0.97 & $2.9^{+4.8}_{-4.7}$ & 0.99 & $0.3^{+1.9}_{-1.8}$ & 0.70 & $11.5^{+8.7}_{-6.0}$ \\
5 & J0140$+$234 & 4.0 & 6.9 & 2.0 & 2.0 & ... & 0.72 & $-2.9^{+6.0}_{-8.0}$ & 0.72 & $4.9^{+14.7}_{-9.2}$ & 0.84 & $3.1^{+2.7}_{-3.0}$  & ... & ... \\
6 & PG~0947$+$396 & 4.0 & 6.0 & 2.0 & 2.0 & 10.5 & 0.96 & $-1.5^{+2.0}_{-2.0}$ & 0.94 & $1.5^{+3.0}_{-4.5}$ & 0.96 & $2.0^{+1.9}_{-1.8}$ & 0.64 & $9.0^{+6.6}_{-5.4}$  \\
7 & J1026$+$523 & 4.9 & 6.0 & 2.0 & 1.9 & 8.4 & 0.95 & $0.3^{+6.3}_{-7.7}$  & 0.95 & $0.7^{+4.9}_{-4.9}$ & 0.98 & $1.3^{+3.1}_{-2.6}$  & 0.94 & $8.7^{+6.2}_{-8.5}$ \\
8 & J1120$+$423 & 4.0 & 6.0 & 2.0 & 2.0 & 5.5 & 0.99 & $1.3^{+3.3}_{-3.0}$ & 0.98 & $2.5^{+3.6}_{-4.4}$ & 0.99 & $4.5^{+1.2}_{-1.9}$ & 0.93 & $7.6^{+2.9}_{-7.9}$ \\
9 & PG~1121$+$422 & 4.0 & 6.0 & 1.9 & 1.9 & 2.1 & 0.99 & $-4.3^{+3.1}_{-3.7}$ & 0.99 & $3.7^{+2.3}_{-2.6}$ & 1.00  & $5.1^{+2.6}_{-2.7}$ & 0.98 & $10.0^{+5.5}_{-3.5}$ \\
10 & PG~1202$+$281 & 4.0 & 5.9 & 2.0 & 2.0 & 2.1 & 0.98 & $3.7^{+2.3}_{-1.8}$ & 0.98 & $3.0^{+2.5}_{-1.5}$ & 0.98 & $1.3^{+1.4}_{-1.3}$ & 0.92 & $7.0^{+3.5}_{-3.0}$ \\
11 & J1217$+$333 & 4.0 & 6.0 & 2.0 & 2.0 & 2.1 & 0.96 & $0.5^{+4.0}_{-2.0}$ & 0.95 & $-0.5^{+8.0}_{-5.0}$ & 0.96 & $1.3^{+2.6}_{-2.9}$ & 0.95 & $8.2^{+4.0}_{-4.6}$ \\
12 & VIII~Zw~218 & 3.9 & 5.9 & 2.9 & 2.0 & 6.0 & 0.98 & $2.5^{+4.0}_{-3.7}$ & 0.97 & $2.6^{+5.0}_{-5.3}$ & 0.98 & $10.1^{+2.0}_{-2.3}$ & 0.97 & $16.0^{+6.5}_{-3.0}$ \\
13 & PG~1322$+$659 & 3.9 & 5.9 & 1.9 & 1.9 & 2.0 & 0.96 & $-5.6^{+1.9}_{-1.7}$ & 0.96 & $0.5^{+3.0}_{-2.5}$ & 0.99 & $0.5^{+1.0}_{-1.0}$ & 0.93 & $8.5^{+5.9}_{-5.7}$ \\
14 & J1415$+$483 & 5.0 & 6.0 & 2.0 & 2.0 & 3.9 & 0.92 & $3.5^{+3.4}_{-2.8}$ & 0.88 & $4.7^{+6.4}_{-6.5}$ & 0.94 & $3.3^{+1.9}_{-2.1}$ & 0.88 & $14.7^{+6.3}_{-6.7}$ \\
15 & PG~1440$+$356 & 4.0 & 5.9 & 2.0 & 2.0 & 4.9 & 0.94 & $-0.8^{+4.9}_{-4.8}$ & 0.92 & $4.2^{+4.1}_{-4.4}$ & 0.97 & $10.3^{+2.2}_{-2.9}$ & 0.94 & $1.5^{+0.7}_{-0.6}$ \\
16 & J1456$+$380 & 3.9 & 5.9 & 1.9 & 1.9 & 2.1 & 0.95 & $0.4^{+4.5}_{-4.4}$ & 0.94 & $7.6^{+6.4}_{-8.1}$ & 0.94 & $3.8^{+2.1}_{-2.2}$ & 0.74 & $11.7^{+4.0}_{-3.8}$ \\
17 & J1526$+$275 & 4.9 & 5.9 & 3.0 & 3.0 & 3.9 & 0.89 & $-3.5^{+4.3}_{-3.8}$ & 0.89 & $-0.9^{+3.3}_{-3.8}$ & 0.95 & $1.5^{+3.0}_{-3.0}$ & 0.92 & $16.6^{+5.0}_{-7.4}$ \\
18 & J1540$+$355 & 4.0 & 5.9 & 1.1 & 1.9 & 1.1 & 0.98 & $-0.5^{+2.0}_{-4.0}$ & 0.99 & $6.7^{+3.5}_{-4.0}$ & 0.99 & $2.0^{+2.3}_{-2.2}$ & 0.96 & $3.5^{+2.0}_{-2.0}$ \\
19 & J1619$+$501 & 4.0 & 5.9 & 1.1 & 1.1 & 1.1 & 0.86 & $-6.1^{+3.8}_{-3.7}$ & 0.77 & $1.0^{+8.0}_{-3.5}$ & 0.92 & $1.6^{+1.8}_{-1.7}$ & 0.60 & $2.5^{+4.6}_{-2.8}$ \\
20 & PG~2349$-$014 & 5.4 & 6.9 & 3.0 & 3.0 & 5.0 & 0.98 & $-2.5^{+6.5}_{-4.3}$ & 0.94 & $4.8^{+10.2}_{-9.7}$ & 0.98 & $2.5^{+2.5}_{-2.4}$ & 0.94 & $21.9^{+4.0}_{-5.6}$ \\

 \hline \\

21 & PG~0003$+$158 & ... & 5.9 & 2.1 & 2.0 & 4.0 & ...  & ... & 0.92  & $5.7^{+8.7}_{-11.1}$ & 0.95 &  $1.7^{+3.6}_{-3.6}$ & 0.88 & $12.4^{+7.8}_{-6.4}$ \\
22 & PG~0043$+$039 & ... & ... & 2.9 & 2.9 & 4.9 & ...  & ... & ... & ... & 0.85 & $3.8^{+5.7}_{-6.9}$ & 0.72 & $12.6^{+8.7}_{-7.9}$ \\
23 & J0848$+$282 & 3.3 & ... & 2.0 & 2.0 & 3.1 & 0.95 & $-1.0^{+4.9}_{-10.3}$ & ... & ... &  0.99 & $2.6^{+3.5}_{-3.0}$ & 0.97 & $10.9^{+5.5}_{-8.6}$ \\
24 & J1059$+$665 & 5.9 & .. & 2.0 & 2.0 & 7.9 & 0.88 & $2.7^{+3.2}_{-3.0}$ & ... & ... & 0.90 & $3.1^{+2.9}_{-2.8}$ & 0.83 & $3.8^{+6.9}_{-6.9}$ \\
25 & PG~1114$+$445 & 3.9 & 12.0 & 2.0 & 2.0 & 3.0 & 0.97 & $1.5^{+10.5}_{-9.4}$ & 0.96 & $8.6^{+7.6}_{-9.5}$ & 0.97 & $2.5^{+1.0}_{-2.0}$ & 0.89 & $11.4^{+9.4}_{-11.2}$ \\
26 & J1152$+$453 & 4.2 & ... & 2.0 & 2.0 & 3.0 & 0.84 & $-0.0^{+4.8}_{-4.1}$ & ... & ... & 0.93 & $0.0^{+6.1}_{-5.6}$ & 0.73 & $2.6^{+5.3}_{-5.7}$ \\
27 & J1203$+$455 & 4.0 & 6.0 & 2.0 & 2.0 & 2.9 & 0.95 & $-6.4^{+4.8}_{-5.5}$ & 0.90 & $1.0^{+6.0}_{-4.2}$ & 0.97 & $1.0^{+1.6}_{-1.5}$ & 0.96 & $11.9^{+6.1}_{-6.7}$ \\
28 & J1207$+$262 & ... & 10.1 & 2.0 & 2.0 & 3.0 & ... & ... & 0.84 & $-0.4^{+10.6}_{-12.9}$ & 0.93 & $8.7^{+3.3}_{-3.6}$ & 0.95 & $11.6^{+6.0}_{-6.5}$ \\
29 & J1220$+$405 & 4.0 & ... & 2.0 & 1.9 & 3.8 & 0.77 & $-2.5^{+3.5}_{-4.0}$ & ... & ... & 0.84 & $3.4^{+2.5}_{-2.7}$ & 0.63 & $4.5^{+8.0}_{-4.5}$ \\
30 & PG~1354$+$213  & ... & 4.9 & 3.0 & 2.9 & 4.9 & ... & ... & 0.88 & $-1.5^{+7.2}_{-6.4}$ & 0.99 & $3.4^{+4.5}_{-4.4}$ & 0.95 & $3.2^{+5.2}_{-5.3}$ \\
31 & J1408$+$630 & 5.3 & 11.6 & 1.9 & 1.8 & 1.8 & 0.96 & $4.8^{+5.1}_{-5.2}$ & 0.98 & $4.4^{+8.4}_{-9.8}$ & 0.98 & $12.1^{+2.7}_{-2.7}$ & 0.96 & $20.6^{+1.8}_{-2.0}$ \\
32 & J1453$+$343 & 4.3 & ... & 1.9 & 1.9 & 4.0 & 0.91 & $-1.8^{+4.7}_{-4.3}$ & ... & ... & 0.97 & $5.3^{+2.7}_{-3.1}$ & 0.96 & $9.0^{+1.5}_{-3.5}$ \\

33 & J1515$+$480 & ... & 8.1 & 1.0 & 1.0 & 1.9 & ... & ... & 0.96 & $-1.8^{+15.5}_{-13.3}$ & 0.99 & $18.0^{+2.6}_{-2.7}$ & 0.98 & $24.9^{+3.4}_{-4.6}$ \\
34 & J1935$+$531 & 4.0 & 5.9 & 1.9 & 1.9 & ... & 0.88 & $-0.3^{+5.5}_{-5.1}$ & 0.85 & $1.5^{+12.5}_{-18.0}$ & 0.97 & $ 0.9^{+2.5}_{-2.5}$ & ... & ... \\
35 & PG~2130$+$099 & 5.1 & 8.0 & 2.1 & 2.9 & 4.9 & 0.91 & $-3.7^{+2.9}_{-2.9}$ & 0.91 & $-2.5^{+5.5}_{-13.0}$ & 0.94 & $6.5^{+2.3}_{-1.7}$ & 0.82 & $3.6^{+3.0}_{-3.3}$ \\
36 & PG~2214$+$139 & 5.9 & 6.0 & 2.0 & 2.0 & 4.1 & 0.78 & $-3.9^{+5.4}_{-6.2}$ & 0.88 & $-4.5^{+6.4}_{-6.8}$ & 0.97 & $2.8^{+2.8}_{-2.8}$ & 0.95 & $15.1^{+6.8}_{-6.7}$ \\
37 & PG~2251$+$113  & ... & ... & 2.1 & 2.0 & 4.5 & ... & ...  & ... & ... & 0.97 & $6.1^{+11.3}_{-11.3}$ & 0.87 & $23.3^{+4.8}_{-8.7}$ \\

 \hline


\end{tabular}

}
\vspace{0.05cm}

\parbox{\linewidth}{
        \vspace{1em} 
        \noindent
        \textbf{Note.} Columns are (1) serial number, (2) object name, (3)--(7) median cadence values in $B$, $V$, $g$, $r$, and $i-$ band light curves, respectively, (8) peak correlation coefficient from ICCF between $g$ and $B-$bands, (9) observed-frame time lag between $g$ and $B-$bands from ICCF,  (10) peak correlation coefficient from ICCF between $g$ and $V-$bands, (11) observed-frame time lag between $g$ and $V-$bands from ICCF, (12) peak correlation coefficient from ICCF between $g$ and $r-$bands, (13) observed-frame time lag between $g$ and $r-$bands from ICCF, (14) peak correlation coefficient from ICCF between $g$ and $i-$bands, and (15) observed-frame time lag between $g$ and $i-$bands from ICCF.}

\end{table*}






\begin{figure}
\centering
\hspace{-0.6cm}
\includegraphics[scale=0.54]{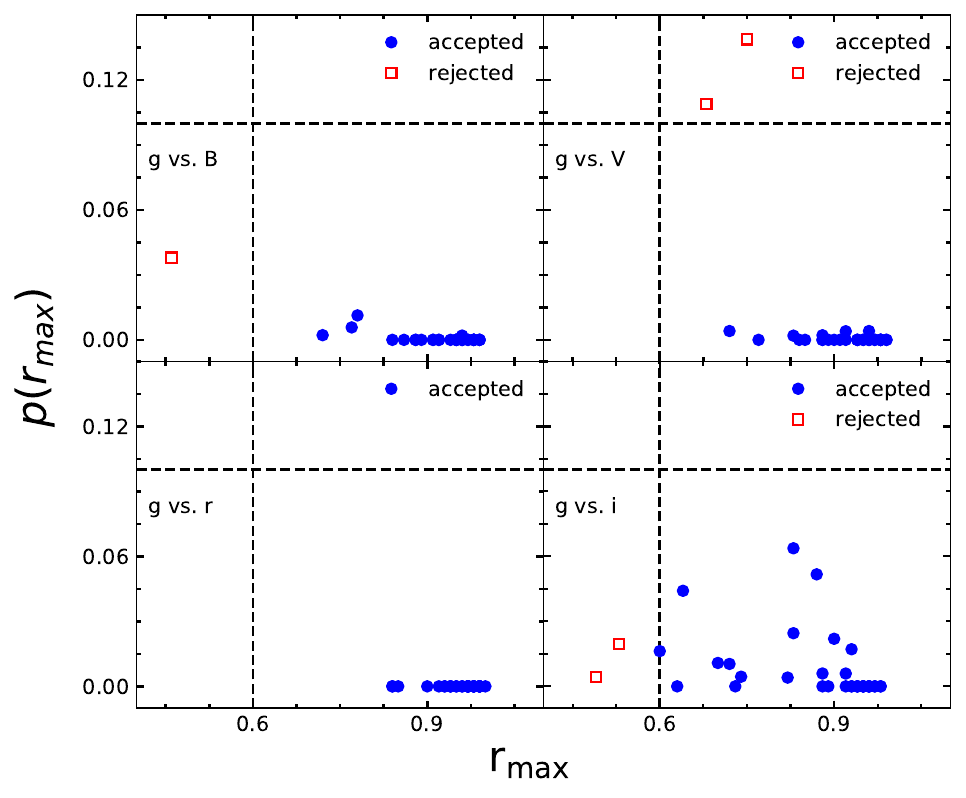}

\caption{The probability that uncorrelated light curves generate a cross-correlation coefficient of at least the maximum cross-correlation coefficient ($r_{max}$), denoted as $p(r_{max})$, is shown as a function of $r_{max}$ for the SAMP sample. In each panel, a vertical dashed line represents $r_{max}$=0.6, while a horizontal dashed line corresponds to $p(r_{max})$=0.1.}
\label{fig:pval}
\end{figure}

\subsubsection{lag quality assessment} \label{sec:lag_qual}

We evaluate the reliability of the lag measurements from the ICCF analysis based on two criteria, following the procedure introduced by \citet{2024ApJ...962...67W}.
First, we check the maximum value of the cross-correlation coefficient $r_{max}$, which indicates the correlation strength between two light curves. Second, we evaluate the probability that the obtained cross-correlation can be produced by two uncorrelated light curves, using the {$\tt PyI^2CCF$}\footnote{\url{https://github.com/legolason/PyIICCF}}. 
For this process we simulate $1,000$ pairs of light curves with an identical sampling as in the observed light curves using damped random walk models. Then, we compute the lag significance indicator, $p(r_{max})_{\tau>0}$, by counting occurrences of positive lags ($\tau \, >$ 0) where the $r_{max}$ value surpasses the observed $r_{max}$ across all simulations. Note that we adopt the same searching window of $\pm 30$ days as used for the lag measurements. 

We consider the lag measurements reliable when $r_{max} \geq 0.6$ and $p(r_{max})_{\tau>0} \leq 0.1$ as similarly adopted by previous studies \citep{2022ApJ...925...52U, 2022ApJ...940...20G, 2024ApJ...962...67W}. 
Except for two targets, all SAMP AGNs satisfy the two criteria as shown in Figure \ref{fig:pval}. 
In addition, we identify five targets, for which the measured lag is consistent with zero within the error between $g$ and $V$ bands as well as $g$ and $r$ bands, suggesting that the lag is not resolved. Thus, we exclude these five AGNs from further analysis and 
obtain reliable inter-band lags measurements for 37 AGNs among the 44 AGNs in the SAMP sample (see Table \ref{tab:samp_laganalysis}).



\begin{figure}
\centering
\hspace{-0.78 cm}
\includegraphics[scale=0.55]{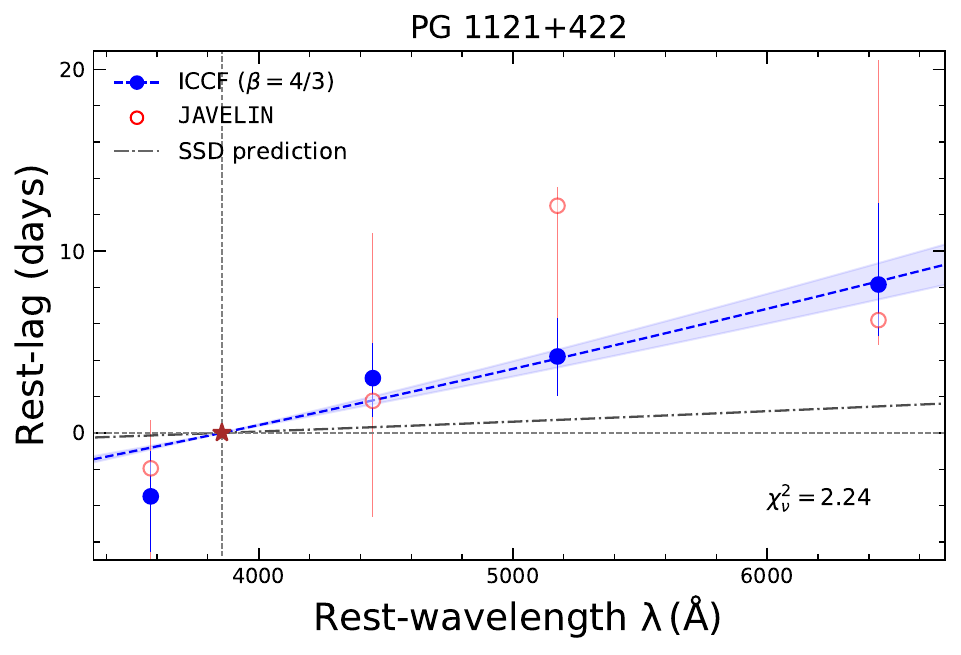}

\caption{Example of the lag-spectrum for PG 1121+422. We present the ICCF lag measurements (blue), and {\tt JAVELIN} lag measurements (open red) in the rest-frame as a function of rest-frame wavelength. The dashed blue line illustrates the best-fitting $\tau \, \propto \, \lambda^{4/3}$ relation to the lags obtained from ICCF with shaded regions indicating the $1 \sigma$ uncertainty from the fitting. Additionally, the black dashed-dotted line shows the lag--wavelength dependence predicted by SSD. The brown star represents the reference point. The vertical and horizontal dotted lines correspond to the rest-frame reference wavelength and zero rest-lag, respectively. The $\chi^{2}$ per degrees of freedom ($\chi_{\nu}^{2}$) is also presented.}
\label{fig:lagspec}
\end{figure}

\begin{figure}
\centering
\includegraphics[scale=0.6]{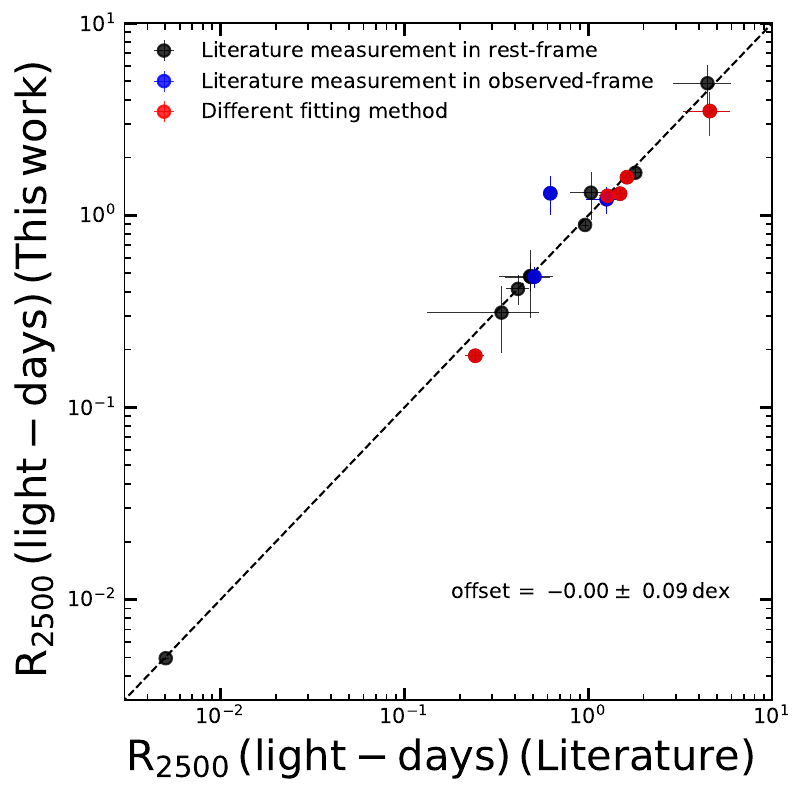}

\caption{Comparison between our measured size of the continuum emitting region at 2500 {\AA}, $R_{2500}$ and the values reported in the literature for the Literature sample. The dashed line represents the 1:1 relation. The literature-measurements (black) are obtained in the rest-frame, while literature-measurements (blue) are acquired in the observed frame. For the literature-measurements (red), the lags were fitted in the lag-wavelength plane with the reference point treated as a free parameter. The figure demonstrates a robust level of consistency between these two sets of measurements, with the mean offset and the rms scatter values explicitly indicated.}
\label{fig:lit_com}
\end{figure}

\subsection{Disk mapping and the size of the continuum emitting region}\label{sec_disk}

\subsubsection{Disk mapping for SAMP sample}

In this section, we present the size of the continuum emitting region (CER) at 2500\AA, R$_{2500}$ based on the disk-mapping analysis with the measured lags.
First, we convert the observed lags to rest-frame values using the $(1+z)$ factor. Second, we fit the lag-spectrum with Equation \ref{fit_eq} 
using the four inter-band lags, i.e., $\tau_{gB}$, $\tau_{gV}$, $\tau_{gr}$, and $\tau_{gi}$ for SAMP AGNs. 
In this process we choose the rest-frame wavelength of the $g-$band at 4722.74 {\AA}/$(1+z)$ as the reference wavelength ($\lambda_0$), for which $\tau$ is set to be zero. 

Consistent with findings from previous studies, the inter-band continuum lags typically scale with wavelength as $\tau \sim \lambda^{4/3}$ \citep{2016ApJ...821...56F, 2017ApJ...836..186J, 2022ApJ...940...20G, 2023ApJ...947...62K}. Therefore, we set the power-law index ($\beta$) to 4/3, with $R_{\lambda_{0}}$ being the only free parameter in the fitting process. 
Then, we compute $R_{2500}$ using the relation $R_{2500} \, = \,  R_{\lambda_{0}}(2500 \mathring{A}/\lambda_{0})^{4/3}$.
Not that we include the lag uncertainties in the fitting process. 
 
In comparison we compute the size of the standard disk at $\lambda = 2500$ {\AA} predicted by \citet{1973A&A....24..337S}, 
presented in units of light-days, 
using the model in \citet{2017ApJ...840...41E}:
\begin{equation} \label{ssd_eq}
R_{\lambda,SSD} = \left(\chi\frac{\kappa \lambda}{hc}\right)^{4/3} \left[\left(\frac{GM_{\text{BH}}}{8 \pi \sigma}\right) \left(\frac{L_{\text{Edd}}}{\eta c^2}\right) (3+\kappa)\lambda_{\text{Edd}}\right]^{1/3}
\end{equation}
where $G$ represents the gravitational constant, $M_{\text{BH}}$ denotes the black hole mass, $\sigma$ is the Stefan--Boltzmann constant, $L_{\text{Edd}}$ corresponds to the Eddington luminosity, and $\lambda_{\text{Edd}}$ signifies the Eddington ratio. $\chi$ is a multiplicative scaling factor that adjusts for systematic discrepancies when converting the annulus temperature, T to wavelength $\lambda$ at a specific characteristic radius. $\eta$ represents the radiative efficiency in converting mass into energy, while $\kappa$ denotes the local ratio of external to internal heating. We assume $\eta$ =0.1 and $\kappa$ =1 for equal heating of the disk by X-rays and viscous effects. The factor $\chi$ is set to 2.49 based on the assumption of a flux-weighted radius \citep{1973A&A....24..337S, 2024ApJ...973..152E}, whereas Wien's law predicts a higher value of $\chi = 4.97$ \citep{2022MNRAS.509.2637N}.
The $R_{2500}$ estimation results for the SAMP sample are provided in Table \ref{tab:disk_size}.

An example of the lag-spectrum and the fitting result is compared to the predicted lag-spectrum based on the SSD in Figure \ref{fig:lagspec} for PG 1121+422 (see also Figure \ref{fig:lagspc1} for a few representative SAMP targets). 
Note that we measure the disk size using the measured lags with the ICCF method as we obtain more conservative errors in the ICCF 
analysis \citep{2022ApJ...940...20G}. Additionally, the ICCF measurements include a cross-correlation reliability test based on the $r_{max}$ and $p(r_{max})_{\tau>0}$ values (see Section \ref{sec:lag_qual}). This approach ensures a robust and reliable measurements of inter-band lag measurements for our investigation.

\subsubsection{Disk mapping for Literature sample}

We performed the disk-mapping analysis for the Literature sample. First we noticed that the $R_{2500}$ estimation is not uniformly determined for Literature sample. For a few AGNs, $R_{2500}$ was measured from the lag-spectrum in the observed-frame \citep{2019ApJ...870..123E, 2020ApJ...896....1C, 2022A&A...659A..13F}, while for other AGNs, the lags were fitted with the reference point as a free parameter in the lag-spectrum, allowing the wavelength of zero time delay to vary from $\lambda_{0}$ \citep{2018ApJ...857...53C, 2021ApJ...922..151K, 2023ApJ...947...62K, 2023ApJ...953..137M, 2023MNRAS.523..545D}. To achieve consistent $R_{2500}$ estimates for Literature sample, we decided to perform the disk fitting procedure as adopted for the SAMP sample. We collected the reported inter-band lags from the original papers, and subsequently performed disk-fitting using Equation \ref{fit_eq}.

For a consistency check, we compare our estimates with that reported in the literature in Figure \ref{fig:lit_com}, which shows a small offset with an rms scatter of 0.09 dex. We present $R_{2500}$ for Literature sample in Table \ref{tab:disk_size} and the fitting results for several additional Literature targets in Figure  \ref{fig:lagspc_lit}.

We find one common target, PG~2130+099 in the SAMP and Literature samples. 
Using the reported lags in the literature, we obtain $R_{2500}$ to be $1.30 \pm 0.30$ light-days, which is consistent within the uncertainty with our estimate of $R_{2500} = 1.8 \pm 1.6$ light-days using the lag measured based on SAMP+ZTF data. In the following analysis we adopt the results from the SAMP sample analysis for PG~2130+099.

\begin{figure}
\centering
\hspace{-1.2cm}
\includegraphics[scale=0.7]{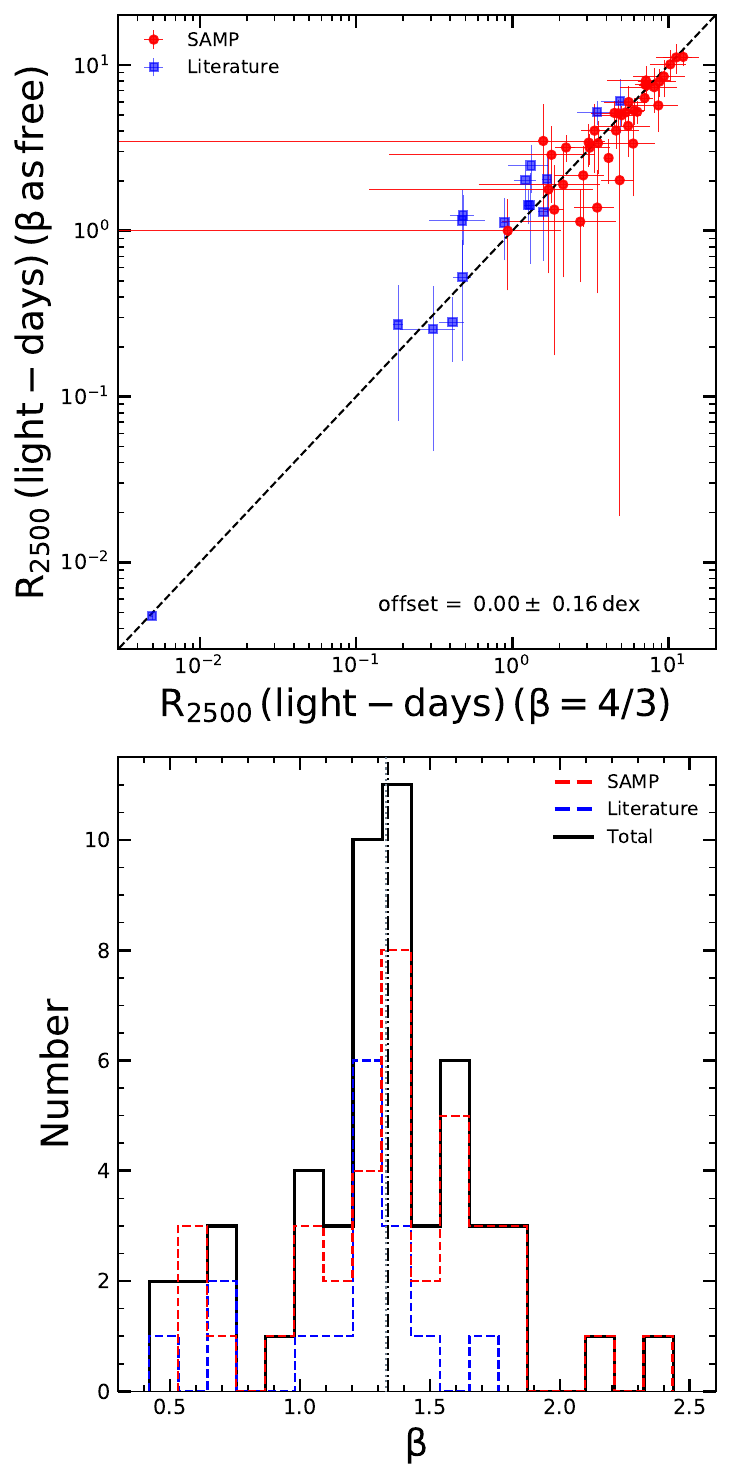}

\caption{ Top: Comparison between $R_{2500}$ values obtained with a fixed $\beta=4/3$ and when $\beta$ is considered as a free parameter for SAMP sample (red) and Literature sample (blue). The dashed black line represents the 1:1 line.
Bottom: Distributions of $\beta$ for SAMP (dashed red line), Literature (dashed blue line), and Total sample (solid black line) are presented. The median $\beta$ value for Total sample is indicated by dashed-dotted black vertical line, while the dotted gray vertical line shows $\beta=4/3$.}
\label{fig:beta_dist}
\end{figure}

\subsubsection{Test with a free parameter $\beta$}

 For a consistency test, we perform the disk-fitting analysis by allowing the power of the wavelength to be free in Equation \ref{fit_eq}. In the top panel of Figure \ref{fig:beta_dist}, we compare $R_{2500}$ values obtained with $\beta$ as a free parameter to those obtained when $\beta$ is fixed to 4/3 for SAMP and Literature samples. Notably, we find similar $R_{2500}$ values, with an rms scatter of 0.16 dex observed between the cases where $\beta$ is treated as a free parameter and when it is fixed to the value of 4/3. However, in a few cases, the fitting results indicate larger  uncertainties than best-fit $\beta$ values exceeding $100\%$. This results in weaker constraints on the recovered parameters $R_{2500}$ and $\beta$ compared to the case when $\beta$ is fixed at 4/3. Therefore, in the bottom panel of Figure \ref{fig:beta_dist}, we display the distributions of $\beta$ for SAMP, Literature and Total sample (SAMP+Literature). We find that for the Total sample, the median value of $\beta$ is approximately $1.34\pm0.39$, which aligns with the expected value of 4/3 from the SSD model. Our obtained $\beta$ value is also consistent with the reported median value of $1.1^{+0.7}_{-0.4}$ ($1.6^{+0.4}_{-0.5}$) for the Parent (Core) sample by \citet{2022ApJ...940...20G}. Therefore, the collective evidence presented in Figure \ref{fig:beta_dist} suggests that the inter-band optical continuum lags broadly follow to the power-law index $\beta=4/3$ with wavelength, as  predicted by the SSD model.

\subsection{Consistency check with Guo+22 sample}\label{guo}

In addition to the SAMP and Literature samples, we incorporate the dataset studied by \citet{2022ApJ...940...20G} to further assess consistency with their results. 
For Guo+22 sample, only 2 data points, $\tau_{gr}$ and $\tau_{gi}$ were used in the lag-spectrum, excluding the reference point, hence, their estimated disk sizes more uncertain. 

To ensure robustness, we consider disk size estimates to be most reliable when the lag-spectrum includes at least three data points (excluding the reference point), which results in 35 targets in SAMP and 16 targets from Literature sample.  Consequently, our subsequent analysis will focus on the most reliable sample comprising SAMP(35) + Literature(16), collectively referred to as Total sample.

For consistency checks, we also incorporate the Guo+22 dataset (84 targets) and two additional targets from the SAMP sample, both of which have only two data points in their lag-spectrum. These supplementary analyses provide additional validation of our findings (see Table \ref{tab:fit}).

\begin{table}
\centering
\movetableright= -30mm
 \caption{Parameters estimated from the lag-spectrum for the AGNs from SAMP and Literature samples}
 \label{tab:disk_size}

\resizebox{8cm}{!}{
\fontsize{20pt}{20pt}\selectfont
\begin{tabular}{clcccc} \hline \hline

Number & Name & $R_{2500}$ & $R_{2500,\text{SSD}}$ & $\chi^2_{\nu}$  & Ref. 
\\ 
 & & (light-days) & (light-days)  & & \\
(1) & (2) & (3) & (4) & (5) & (6)
\\ \hline

&    & SAMP sample &  &  &  \\

1 & Mrk~1501 & $3.1\pm0.1$ & 0.36 & 14.72  & ... \\
2 & PG~0026$+$129 & $8.6\pm2.9$ & 0.61 & 0.27 & ...  \\
3 & PG~0052$+$251 & $7.2\pm0.8$ & 1.01 & 2.82  & ...  \\
4 & J0101$+$422 & $5.3\pm0.5$ & 1.22 & 7.07  & ...  \\
5 & J0140$+$234 & $11.2\pm2.8$ & 1.27 & 3.37  & ... \\
6 & PG~0947$+$396 & $4.1\pm0.2$ & 0.79 & 6.54 & ... \\
7 & J1026$+$523 & $3.5\pm1.0$ & 0.45 & 0.50  & ... \\
8 & J1120$+$423 & $3.5\pm0.4$ & 0.78 & 2.31  & ... \\
9 & PG~1121$+$422 & $4.8\pm0.6$ & 0.83 & 2.24  & ... \\
10 & PG~1202$+$281 & $2.8\pm1.0$ & 0.63 & 1.35  & ... \\
11 & J1217$+$333 & $2.7\pm1.8$ & 0.44 & 0.20  & ... \\
12 & VIII~Zw~218 & $7.1\pm1.0$ & 0.71 & 1.11  & ... \\
13 & PG~1322$+$659 & $3.4\pm0.2$ & 0.82 & 21.45  & ... \\
14 & J1415$+$483 & $7.0\pm0.9$ & 0.65 & 2.88  & ... \\
15 & PG~1440$+$356 & $8.1\pm3.2$ & 0.48 & 1.39 & ... \\
16 & J1456$+$380 & $5.5\pm1.7$ & 1.21 & 0.62  & ... \\
17 & J1526$+$275 & $6.3\pm0.7$ & 1.29 & 6.57  & ... \\
18 & J1540$+$355 & $2.1\pm1.5$ & 0.51 & 0.82  & ... \\
19 & J1619$+$501 & $1.6\pm2.3$ & 0.71 & 0.36  & ... \\
20 & PG~2349$-$014 & $8.7\pm2.4$ & 1.30 & 0.53  & ... \\

\hline \\
21 & PG~0003$+$158 & $7.1\pm3.1$ & 3.35 & 0.40  & ... \\
22 & PG~0043$+$039 & $6.0\pm1.0$ & 2.00 & 0.87  & ... \\
23 & J0848$+$282 & $5.0\pm1.6$ & 0.55 & 0.23  & ... \\
24 & J1059$+$665 & $1.8\pm0.3$ & 1.15 & 12.19  & ... \\
25 & PG~1114$+$445 & $4.5\pm1.3$ & 0.82 & 0.42  & ... \\
26 & J1152$+$453 & $0.9\pm1.1$ & 0.82 & 0.09  & ... \\
27 & J1203$+$455 & $4.9\pm1.2$ & 1.50 & 2.15 & ... \\
28 & J1207$+$262 & $5.9\pm2.2$ & 0.67 & 0.30  & ... \\
29 & J1220$+$405 & $2.2\pm0.3$ & 0.65 & 4.67  & ... \\
30 & PG~1354$+$213 & $1.7\pm1.6$ & 1.26 & 0.28  & ... \\
31 & J1408$+$630 & $9.3\pm3.4$ & 1.01 & 0.27  & ... \\
32 & J1453$+$343 & $4.6\pm1.8$ & 0.76 & 0.04  & ... \\
33 & J1515$+$480 & $12.4\pm3.2$ & 1.44 & 0.71  & ... \\
34 & J1935$+$531 & $3.1\pm1.1$ & 1.68 & 1.75  & ... \\
35 & PG~2130$+$099 & $1.8\pm1.6$ & 0.43 & 0.48  & ... \\
36 & PG~2214$+$139 & $5.5\pm2.1$ & 0.75 & 0.90  & ... \\
37 & PG~2251$+$113 & $10.2\pm2.7$ & 2.30 & 0.60  & ... \\

 \hline 
 \hline
 &   & Literature sample &  &  &  \\

38 & Ark~120  & $1.31\pm0.37$ & 0.42 & 0.77  & 1  \\
39 & Fairall~9 & $1.67\pm0.12$ & 0.55 & 0.67  & 2 \\
40 & MCG~$+08$ &  &  &  &   \\
& $-11-011$ & $0.48\pm0.19$ & 0.16 & 1.24  & 3 \\
41 & Mrk~110 & $0.41\pm0.07$ & 0.20 & 0.54  & 4 \\
42 & Mrk~142 & $0.48\pm0.06$ & 0.08 & 0.44  & 5 \\
43 & Mrk~509 & $1.21\pm0.19$ & 0.53 & 0.48  & 6 \\
44 & Mrk~817 & $1.30\pm 0.10$ & 0.25 & 0.48  & 7 \\
45 & NGC~2617 & $0.31\pm 0.12$ & 0.10 & 0.39  & 3 \\
46 & NGC~4151 & $0.48\pm 0.09$ & 0.07 & 0.37  & 8 \\
47 & NGC~4593 & $0.19\pm 0.01$ & 0.06 & 0.85  & 9 \\

48 & NGC~5548 & $0.89\pm 0.05$ & 0.19 & 0.77  & 10 \\
49 & Mrk~335 & $1.58\pm 0.12$ & 0.21 & 0.82  & 11 \\
50 & IRAS~09149$-$6206 & $4.87\pm 1.20$ & 0.64 & 0.53  & 12 \\
51 & Mrk~876 & $3.49\pm 0.89$ & 1.04 & 0.98  & 13 \\
52 & PG~1119+120 & $1.26\pm 0.04$ & 0.21 & 3.20  & 14 \\
53 & NGC~4395 & $0.005\pm<0.001$ & 0.001 & 1.16  & 15 \\
54 & PG~2130+099 & $1.30\pm 0.30$ & 0.43 & 0.93  & 16 \\

\hline


\end{tabular}
}

\parbox{\linewidth}{
        \vspace{1em} 
        \noindent
        \textbf{Note.} Columns are (1) serial number, (2) object name, (3) size of the continuum emitting region at 2500 {\AA} derived from the fits to the lag-spectrum, (4) SSD predicted disk size at 2500 {\AA}, (5) $\chi^2$/degrees of freedom, and (6) Ref. (reference for the continuum inter-band lag measurements used to obtain $R_{2500}$): 1 \citet{2020MNRAS.494.1165L}, 2 \citet{2020MNRAS.498.5399H}, 3 \citet{2018ApJ...854..107F}, 4 \citet{2021MNRAS.504.4337V}, 5 \citet{2020ApJ...896....1C}, 6 \citet{2019ApJ...870..123E}, 7 \citet{2021ApJ...922..151K}, 8 \citet{2017ApJ...840...41E}, 9 \citet{2018ApJ...857...53C}, 10 \citet{2016ApJ...821...56F}, 11 \citet{2023ApJ...947...62K}, 12 \citet{2023MNRAS.525.4524G}, 13 \citet{2023ApJ...953..137M}, 14 \citet{2023MNRAS.523..545D}, 15 \citet{2022ApJ...934L..37M}, and 16 \citet{2022A&A...659A..13F}. 
    }



\end{table}

\section{Results and Discussion}
\label{sec:result}
\subsection{Disk Size Anomalies: larger than expected}

\begin{figure}
\centering
\includegraphics[scale=0.7]{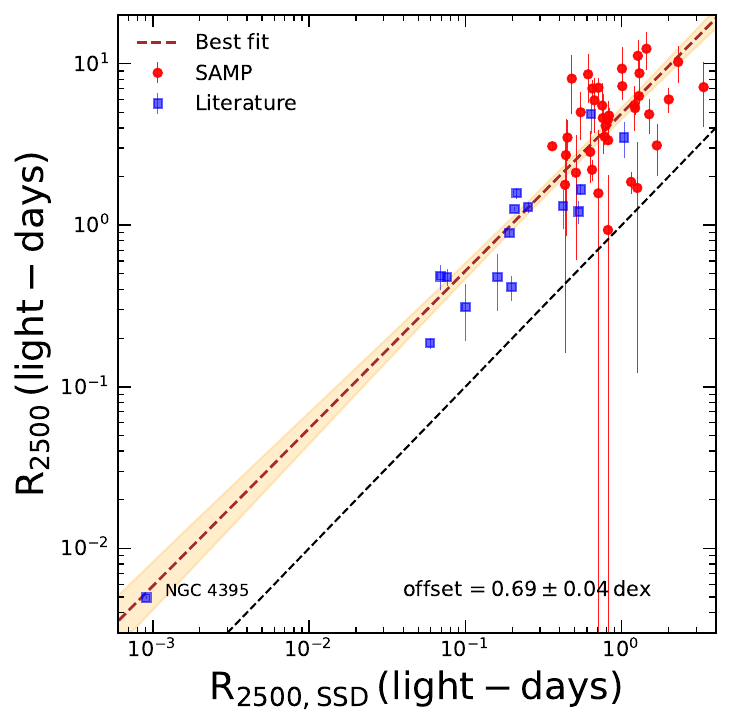}

\caption{Comparison between $R_{2500}$ derived from lag-spectrum fitting and that expected from the SSD model ($R_{2500, \text{SSD}}$) is shown for the SAMP sample (red), and Literature sample (blue). The best-fit is presented by the dashed brown line, whereas the orange-shaded region indicates the range driven by the uncertainties on the best-fit. The 1:1 relation is shown by the dashed black line.  The  offset value with 1$\sigma$ uncertainty from the best-fit relation is also indicated.}
\label{fig:comp_obsd}
\end{figure}

\begin{figure*}
\centering
\includegraphics[scale=0.6]{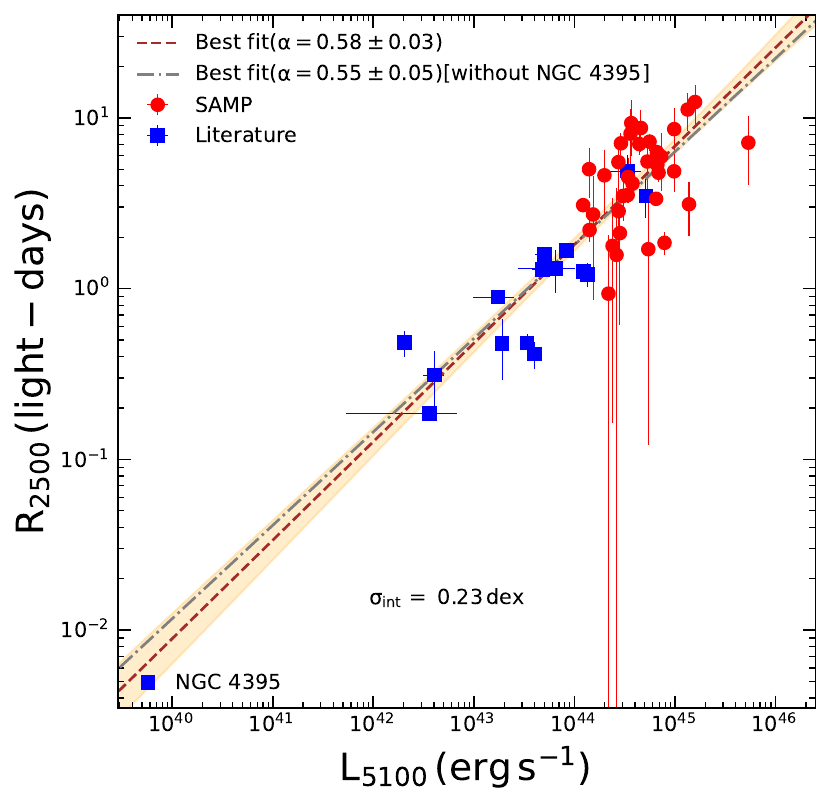}
\includegraphics[scale=0.6]{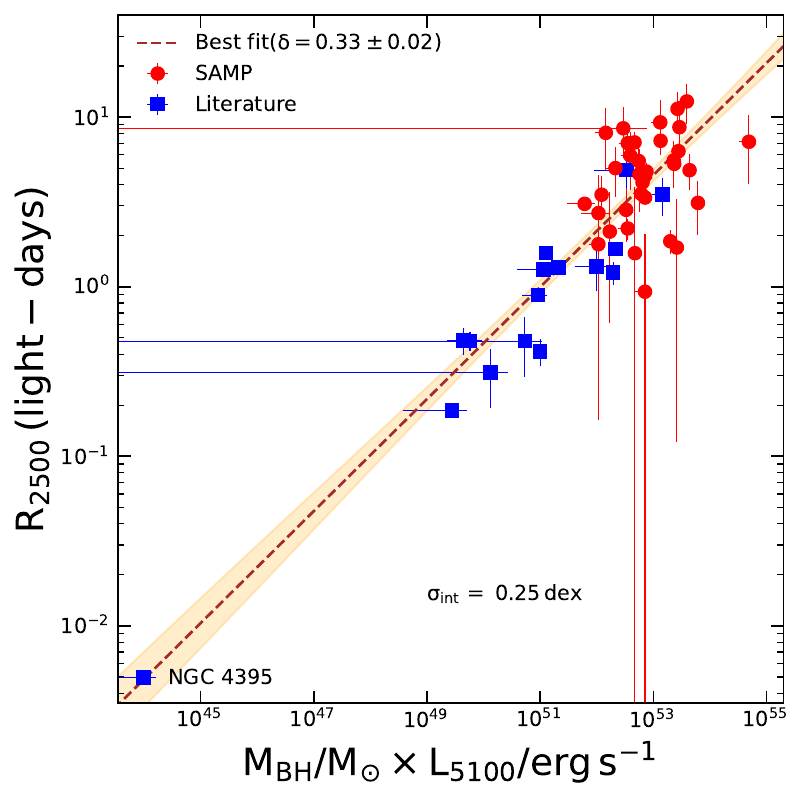}

\caption{Left: The correlation between $R_{2500}$ and the continuum luminosity at 5100 {\AA} is shown for the SAMP sample (red), and Literature sample (blue). The dashed brown line represents the best-fit and the orange-scale region is the range driven by the uncertainties on the best-fit. The dashed gray line represents the best-fit without NGC 4395.  Right: We present the correlation between $R_{2500}$ and $M_{\text{BH}} \times L_{5100}$ for SAMP sample (red), and Literature sample (blue). The best-fit is shown by the dashed brown line with uncertainties indicated by orange-scale region. The intrinsic scatter is also shown at each panel in the figure.}
\label{fig:lag_lm}
\end{figure*}

 
  In this section, we examine the discrepancies observed between the $R_{2500}$ values derived from the lag-spectrum fitting and those predicted by the SSD model ($R_{2500, \text{SSD}}$). In Figure \ref{fig:comp_obsd}, we compare the $R_{2500}$ values with $R_{2500, \text{SSD}}$ assuming $\chi = 2.49$ for the SAMP and Literature samples. We find that $R_{2500}$ linearly correlates with $R_{2500, \text{SSD}}$ with the best-fit slope of $0.98 \pm 0.06$. However, $R_{2500}$ is on average larger than $R_{2500, \text{SSD}}$ by a factor of $\sim5$ ($0.69 \pm 0.04$ dex).  
This trend broadly aligns with findings from previous disk RM campaigns focused on individual nearby type 1 AGNs, which showed a larger $R_{2500}$ from disk-mapping than the SSD-based disk size \citep[i.e.,][]{2016ApJ...821...56F, 1999MNRAS.302L..24C}. Recently, \citet{2024ApJ...973..152E} found that the disk size in Fairall 9 is about 4.5 times larger than the SSD prediction, consistent with our estimated disk size anomalies under the flux-weighted assumption with $\chi = 2.49$. In addition,
\citet{2023ApJ...947...62K} reported that the disk size is approximately 6--11 times larger than predicted by the SSD model in the case of Mrk 335. By collecting disk-mapped AGNs, \citet{2022ApJ...940...20G} presented that the disk size is a factor of $\sim3$ larger than the SSD-based disk size. Additionally, from a microlensing study on 11 gravitationally lensed quasars, \citet{2010ApJ...712.1129M} also found a similar discrepancy in the accretion disk sizes of about $\sim$ 4 times larger than expected from the thin disk model. In contrast, other studies with more luminous and distant AGNs claimed more consistent disk sizes between disk-mapping and the SSD model \citep[i.e.,][]{2019ApJ...880..126H, 2020ApJS..246...16Y}. This discrepancy could potentially be attributed to factors such as lower data quality with lower cadence or there could be a potential dependency of lags on luminosity \citep{2021ApJ...912L..29L, 2022ApJ...940...20G}.

For a consistency check, we repeat the comparison between $R_{2500}$ and $R_{2500, \text{SSD}}$ by including the sample of AGNs from \citet{2022ApJ...940...20G}, who reported an average factor of $\sim$3 larger disk size than $R_{2500, \text{SSD}}$. We notice that Guo+22 sample is located slightly below the best-fit obtained based on our Total sample (SAMP and Literature). We determine the best-fit slope $0.78 \pm 0.05$ and an offset $0.51 \pm 0.03$ dex, implying that there is a luminosity dependency, i.e., higher luminosity AGNs have smaller offset between $R_{2500}$ and $R_{2500, \text{SSD}}$ than lower luminosity AGNs and that the disk size is larger than predicted by the SSD model by a factor of $\sim$3. Note that \citet{2022ApJ...940...20G} performed the disk-mapping using only two inter-band lags. Thus, it is possible that the disk size estimated from the lag-spectrum fitting may have a large uncertainty.



A suggested resolution to the 'larger disk size' issue posits that the measured lags do not solely arise from light travel times between distinct segments in the accretion disk. Instead, there is a plausible contribution from diffuse continuum emission originating from the BLR, attributed to free-free and free-bound hydrogen transitions \citep{2001ApJ...553..695K, 2019MNRAS.489.5284K}. The light travel time from the inner regions of the BLR is significantly extended. Consequently, even a slight involvement of diffuse continuum emission in the observed optical bands will lead to measured lags surpassing the expected delays based solely on a pure accretion disk.  On the other hand, as described by \citet{2021ApJ...907...20K, 2021MNRAS.503.4163K}, an increase in disk size at a particular wavelength may occur due to a substantial height of the X-ray corona. In this context, it is postulated that the X-ray emission is sufficiently strong to induce variability in the UV/optical range.

\subsection{Excess of lags in each band}

 Compared to the best-fit disk model, we notice significant excess rest-frame lag at the $B-$band for 9 objects: PG~0026$+$129 (2.0), J0101$+$422 (2.1), J1120$+$423 (2.8), PG~1202$+$281 (7.1), VIII~Zw~218 (2.6), J1415$+$483 (3.4), J1059$+$665 (8.2), PG~1114$+$445 (2.6), and J1408$+$630 (3.5), where the values in the parenthesis are  the relative residual factors that quantify the difference between the observed lag values and the values predicted by the best-fit model using Equation \ref{fit_eq}, normalized by the best-fit absolute value. Additionally, we found excess rest-frame lag  at the $V-$band for 2 objects: J1456$+$380 (2.0), and J1540$+$355 (5.8), at the $r-$band for 1 object: PG~2130$+$099 (2.3).
 The excess lag observed at the $B-$band can be due to the presence of the Balmer jump at a rest-frame wavelength of 3646 {\AA}. Contamination of this kind has been noted in the lag-spectrum of various objects \citep[i.e., NGC 4593;] []{2018ApJ...857...53C}, however, it is not a prevalent occurrence in all AGNs \citep[i.e., Mrk 817;] []{2021ApJ...922..151K}. Moreover, the excess lags observed at the $V-$band (containing the H$\beta$ line) and $r-$band (containing the H$\alpha$ line) primarily result from contamination by the broad emission lines from BLR.
 We also observe negative rest-frame lags in 9 targets at the $V-$band. However, given the substantial uncertainties associated with these cases, they appear to be consistent with zero lags. It is worth noting that while the cadence of the $V-$ band light curves used may be insufficient to resolve the lag in these cases, our simulation in Section \ref{sec:sim} predicts successful lag recovery within the cadence range of our data, provided the driving continuum light curve ($g-$ band) is densely sampled. Importantly, whether we include or exclude these excess lags (also the negative lags in $V-$band) in the fitting of the lag-spectrum as a function of wavelength, the results remain comparable within the uncertainties. Hence, we opted to incorporate all the reliably measured lags in our fits of the lag-spectrum. We discuss the impact of the excess lag observed in the $i-$band on the fitting of the lag-spectrum separately in Section \ref{sec:lgsp}.

\subsection{Correlation of the disk size with AGN luminosity}



In this section, we investigate the correlation between the size of the disk ($R_{2500}$) with AGN luminosity $L_{5100}$. Left panel of Figure \ref{fig:lag_lm} displays the relationship between $R_{2500}$ and $L_{5100}$ for the SAMP and Literature samples. Notably, a clear positive correlation is evident between $R_{2500}$ and $L_{5100}$. To characterize the $R_{2500}$--$L_{5100}$ relationship, we employ the following linear equation:

\begin{equation}
    \text{log}(R_{2500}/1 \, \text{light-day}) = \zeta \, + \,  \alpha \, \text{log}\left(\frac{L_{5100}}{10^{44.5} \, erg \, s^{-1}}\right)
    \label{eq_rL}
\end{equation}
We utilize the {\tt LINMIX$\textunderscore$ERR} technique \citep{2007ApJ...665.1489K}, employing Bayesian linear regression. This method is proficient at managing intrinsic scatter within the relationship, accommodating errors in measurements for both the independent and dependent variables, and effectively handling correlations among the measurement errors.

The best-fit yields a slope of $\alpha = 0.58 \pm 0.03$ ($0.55 \pm 0.05$) for the SAMP and Literature samples (excluding NGC 4395), with an intrinsic scatter of 0.23 dex.

Our obtained slope aligns well with the reported value of $\alpha = 0.56^{+0.05}_{-0.04}$ for the continuum lag--$L_{5100}$ relation, as observed in nine local AGNs ($z<0.047$) by \citet{2022ApJ...934L..37M}. However, when including NGC 4395 in our analysis, we observe a steeper slope of $\alpha = 0.58$ compared to the expected value of 0.5 from the photo-ionization model of the BLR \citep{1995ApJ...455L.119B}. Notably, the slope becomes marginally consistent (within error) with 0.5 if NGC 4395 is excluded from the fitting. The radiation pressure confined (RPC) cloud model of \citet{2022MNRAS.509.2637N} also predicts a slope of $\alpha = 0.5$ in the $R_{2500}$--$L_{5100}$ relation, indicating that diffuse continuum emission from the BLR significantly or even predominantly contributes to both the optical variability amplitude and the continuum inter-band lags. Coupled with the observed larger-than-expected disk sizes, these findings provide further support for the significant contribution of BLR diffuse continuum emission to optical variability.

However, according to the SSD model, for a fixed $\lambda$, the size of the disk ($R_{2500}$) is proportional to $(M_{\text{BH}} \times L_{5100})^{1/3}$ (see Equation \ref{ssd_eq}). To explore this, we perform a power-law fit for the relation between $R_{2500}$ and $M_{\text{BH}} \times L_{5100}$, as shown in the right panel of Figure \ref{fig:lag_lm}. The best-fit result, $R_{2500} \, \propto \, (M_{\text{BH}} \times L_{5100})^{0.33\pm0.02}$, is in excellent agreement with the SSD predicted value. Moreover, through a linear fit between $L_{5100}$ and $M_{\text{BH}}$, we find $L_{5100} \, \propto \, M_{\text{BH}}^{1.11 \pm 0.10}$ within our sample ($-2.2 < \text{log}\lambda_{\text{Edd}} < 0.4$), with a scatter of 0.83 dex. If the BLR diffuse continuum emission significantly contributes to the continuum lags, $R_{2500}$ would primarily be driven by $L_{5100}$. Consequently, by combining $R_{2500} \, \propto \, L_{5100}^{0.58}$ with $L_{5100} \, \propto \, M_{\text{BH}}^{1.11}$, we obtain $R_{2500} \, \propto \, (M_{\text{BH}} \times L_{5100})^{0.31}$, which is consistent with both the observed relation, $R_{2500} \, \propto \, (M_{\text{BH}} \times L_{5100})^{0.33\pm0.02}$, and the SSD prediction, especially when considering the significant scatter observed between $L_{5100}$ and $M_{\text{BH}}$.

Therefore, these observations collectively suggest that the continuum lags arise from both disk reprocessing and BLR diffuse continuum emission. It is important to note that our sample primarily consists of high-luminosity AGNs, significantly populating the high-luminosity end of the existing $R_{2500}$--$L_{5100}$ relation.

As a consistency check, we include Guo+22 sample to SAMP (37) and Literature (16) samples to study the $R_{2500}$--$L_{5100}$ relation. We find significantly shallower slope of $\alpha = 0.44 \pm 0.03$ ($0.39 \pm 0.04$, without NGC 4395), accompanied by an intrinsic scatter of 0.31 dex (0.30 dex). \citet{2022ApJ...940...20G} reported a similar slope of $\alpha = 0.48 \pm 0.03$ for both the Core and Local sample, after excluding NGC 4395, and $\alpha = 0.31 \pm 0.04$ when considering the Parent sample. However, the shallower slope observed in this case can be because of the underestimation of disk sizes using only two inter-band lags in the lag-spectrum for Guo+22 sample as discussed in Section \ref{guo}.


\begin{figure}
\centering
\includegraphics[scale=0.6]{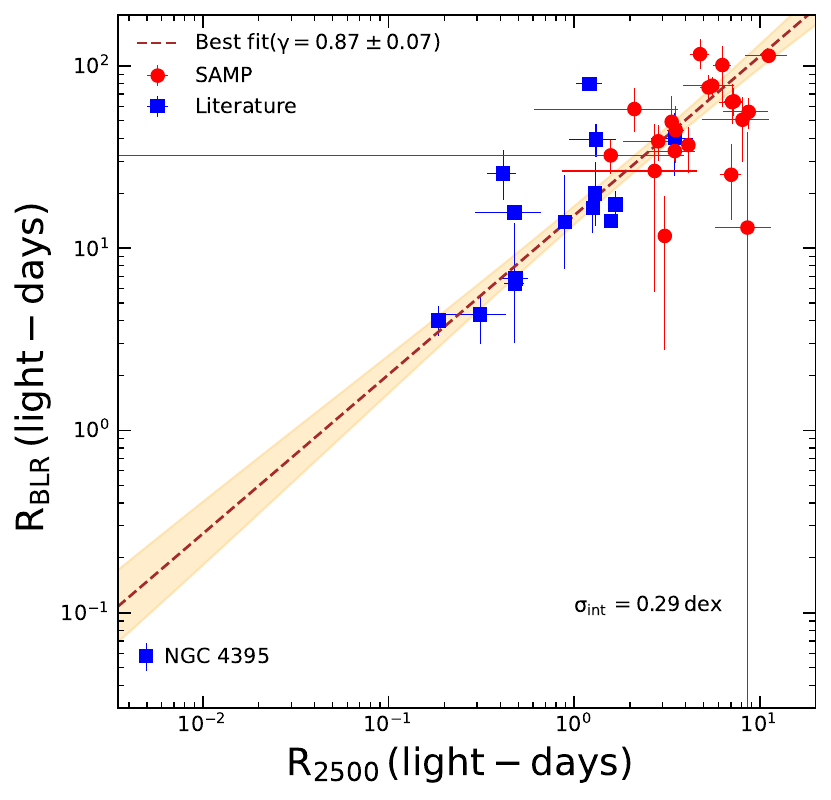}

\caption{BLR size--$R_{2500}$ correlation is shown for the SAMP sample (red), and Literature sample (blue). The best-fit is shown by the dashed brown line, while the orange-scale region indicates the range driven by the uncertainties on the best-fit. The intrinsic scatter is also mentioned in the figure.}
\label{fig:blr_cer}
\end{figure}

\subsection{BLR Size--$R_{2500}$ relation}

Our study reveals that $R_{2500}$ is predominantly influenced by $L_{5100}$, leading to a strong correlation between the two, characterized by the relationship $R_{2500}$ $\propto$ $L_{5100}^{0.58}$. On the other hand,  spectroscopic RM revealed that the BLR size ($R_{\text{BLR}}$) based on the H$\beta$ lag (except for NGC 4395, which has H$\alpha$ lag) is dependent on $L_{5100}$, following a relationship of $R_{\text{BLR}}$ $\propto$ $L_{5100}^{0.402}$ \citep[][]{2024ApJ...962...67W, 2024ApJS..275...13W}. The analogous dependency of $R_{2500}$ and $R_{\text{BLR}}$ on luminosity suggests a substantial contribution from the BLR diffuse continuum emission in the optical continuum bands \citep{2021ApJ...912L..29L, 2022MNRAS.509.2637N, 2022ApJ...940...20G}.

We find 20 targets which have reliable $R_{\text{BLR}}$ and BH mass measurements based on H$\beta$ line from SAMP, and all targets except one (IRAS~09149$-$6206) in Literature sample have $R_{\text{BLR}}$ measurements from H$\beta$--RM. Therefore, by incorporating our SAMP sample along with the sample from Literature, we provide a direct comparison between $R_{\text{BLR}}$ and $R_{2500}$ in Figure \ref{fig:blr_cer} for the most reliable subset. Significantly, there is a noticeable positive correlation between $R_{\text{BLR}}$ and $R_{2500}$. Therefore, we fit the relation using Equation \ref{eq_blr} given below:

\begin{figure}
\centering
\hspace{-1.2cm}
\includegraphics[scale=0.6]{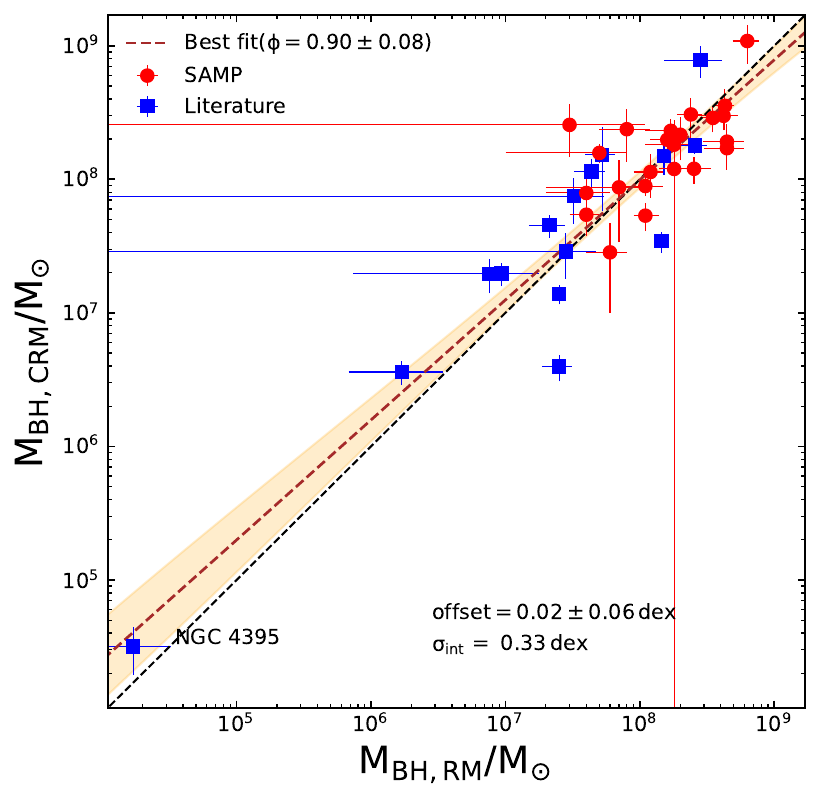}


\caption{Comparison of the continuum-RM based and H$\beta$ RM-based $M_{\text{BH}}$ for the SAMP sample (red), and the Literature sample (blue). The best-fit is presented (dashed brown line) along with the uncertainty range (orange-shaded region). The dashed black line denotes the 1:1 relation. The offset and intrinsic scatter are also provided.}

\label{fig:SE_bh}
\end{figure}

\begin{figure*}
\centering
\hspace{-1.2cm}
\includegraphics[scale=0.5]{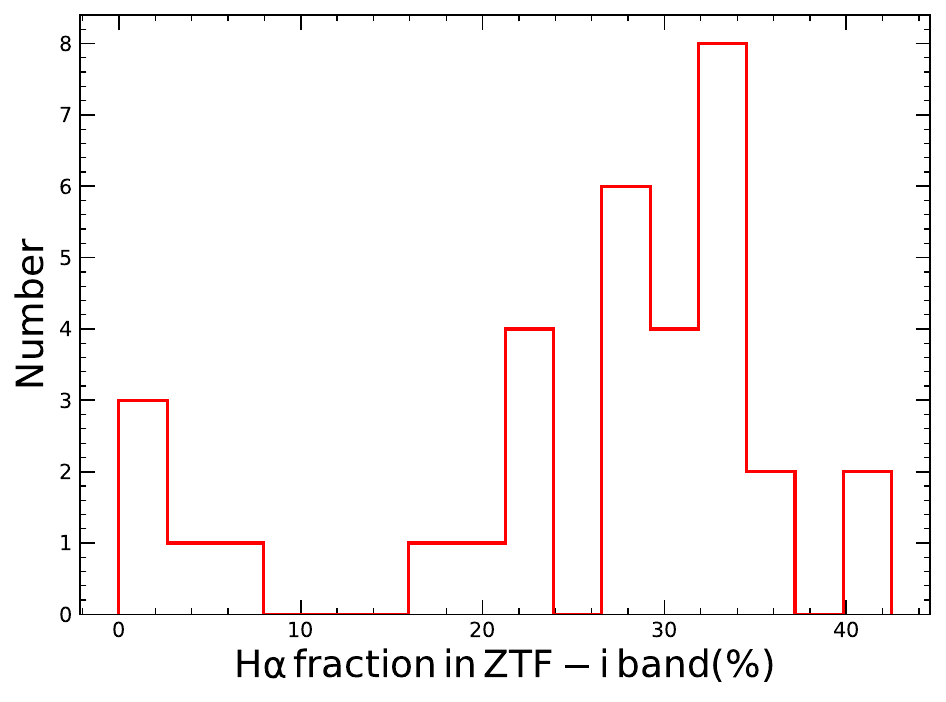}
\includegraphics[scale=0.5]{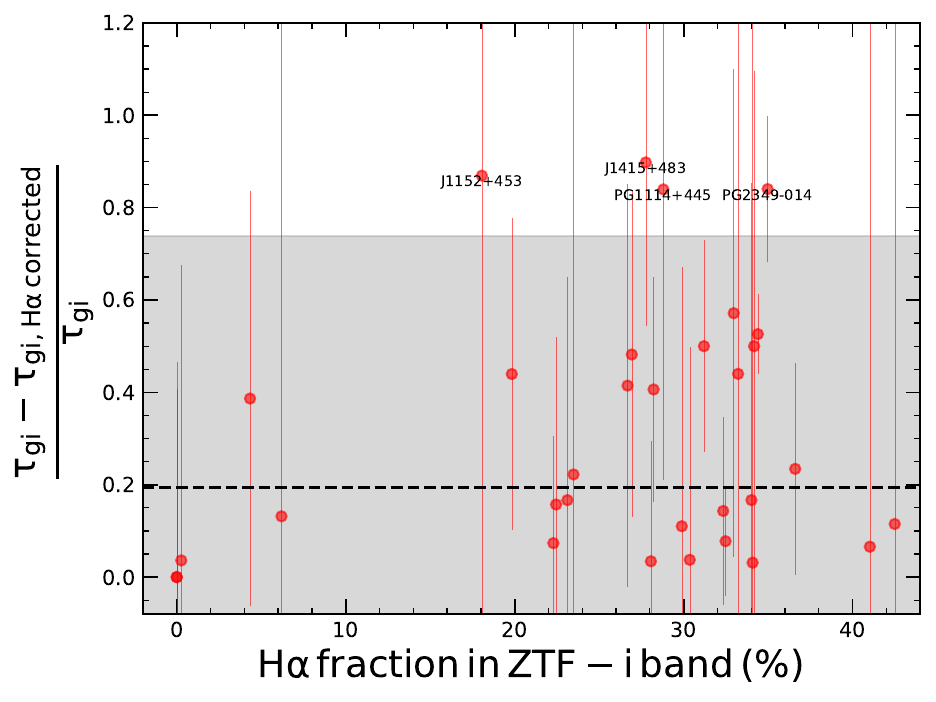}

\caption{ Left: Distribution of the fraction of H$\alpha$ flux (in \%) in the ZTF$-i$ band for the SAMP sample. Right: Excess lags in the $i-$band, before ($\tau_{gi}$) and after correcting for H$\alpha$ contamination ($\tau_{gi, \mathrm{H\alpha \, corrected}}$), plotted as a function of the fraction of H$\alpha$ flux in the $i-$band. The mean value is denoted by dashed black line and the grey-shaded region represents the 2$\sigma$ range of the excess lags. The objects with excess lags exceeding the 2$\sigma$ range are marked separately.}
\label{fig:halp_dist}
\end{figure*}

\begin{equation}
    \text{log}(R_{\text{BLR}}/1 \, \text{light-day}) = K \, + \,  \gamma \, \text{log}(R_{2500}/10^{0.5} \, \text{light-day})
    \label{eq_blr}
\end{equation}
where, $\gamma$ and $K$ represent the slope and intercept, respectively. The reference point is close to the median $R_{2500}$ value within the sample.

Note that by including new SAMP targets, which are the representative of high luminosity AGNs, the sample size is substantially increased. 
To assess the impact of NGC 4395, 
we conduct the analysis with and without including NGC 4395. We find the best-fit slope $\gamma = 0.87\pm0.07$ when NGC 4395 is included.  In contrast, the slope becomes flatter with $\gamma = 0.69\pm0.10$ in the absence of NGC 4395. The slope obtained with NGC 4395 remains consistent with a linear relation ($\gamma = 1.0$) within a 2$\sigma$ uncertainty range, while it deviates significantly from the linear relation when NGC 4395 is excluded. 
Moreover, our derived slope remains consistent with the reported value of $0.98 \pm 0.10$ by \citet{2023ApJ...948L..23W}, which is based on 21 objects and falls within the $1\sigma$ uncertainty when NGC 4395 is included. Hence, we designate the fitting result with NGC 4395 as the fiducial relation due to its inclusion, providing a significantly larger dynamical range and thereby offering better constraints on the slope. 

As an additional consistency check, we extend this comparison by incorporating six additional targets with $R_{\text{BLR}}$ measurements from Guo+22, resulting in similar fitting outcomes. The linear regression fitting results are presented in Table \ref{tab:fit}.

The sublinear relationship observed in our $R_{\text{BLR}}$ -- $R_{2500}$ relation could be attributed to the more compact structures found in high-luminosity AGNs. This trend was previously noted in the BLR \citep{2024ApJ...962...67W, 2024ApJS..275...13W} and torus \citep{2024ApJ...968...59M}, and it likely applies to BLR distributions as well. Consequently, for high-luminosity AGNs with larger $R_{2500}$ values (since $R_{2500} \, \propto \, L_{5100}^{0.58}$), $R_{\text{BLR}}$ would be reduced relative to the value expected from a linear relation.

 Using the best-fit $R_{\text{BLR}}$ -- $R_{2500}$ relation for the fiducial case, the median size ratio $R_{\text{BLR}}$/$R_{2500}$ over the entire $R_{2500}$ range used in this work is found to be $\sim$ $1.25\pm0.12$ dex, suggesting a BLR size roughly 17.8 times larger than $R_{2500}$, a value close to the value of 20.9 (1.32 dex) reported by \citet{2023ApJ...948L..23W}. Furthermore, according to the radiation pressure confined cloud model, \citet{2022MNRAS.509.2637N} predicted that $R_{\text{BLR}}$/$R_{5100}$ would be approximately 8.7 (0.94 dex), which corresponds to $R_{\text{BLR}}$/$R_{2500}$ $\sim$ 22.5 (1.35 dex). Our finding aligns with this predicted value within the $1\sigma$ range. Note that the assumption of radiation pressure confined cloud model is based on the scenario where continuum lags are dominated by the diffuse continuum emission component from BLR with a covering factor of about 0.2.

\begin{figure}
\centering
\hspace{-1.2cm}
\includegraphics[scale=0.65]{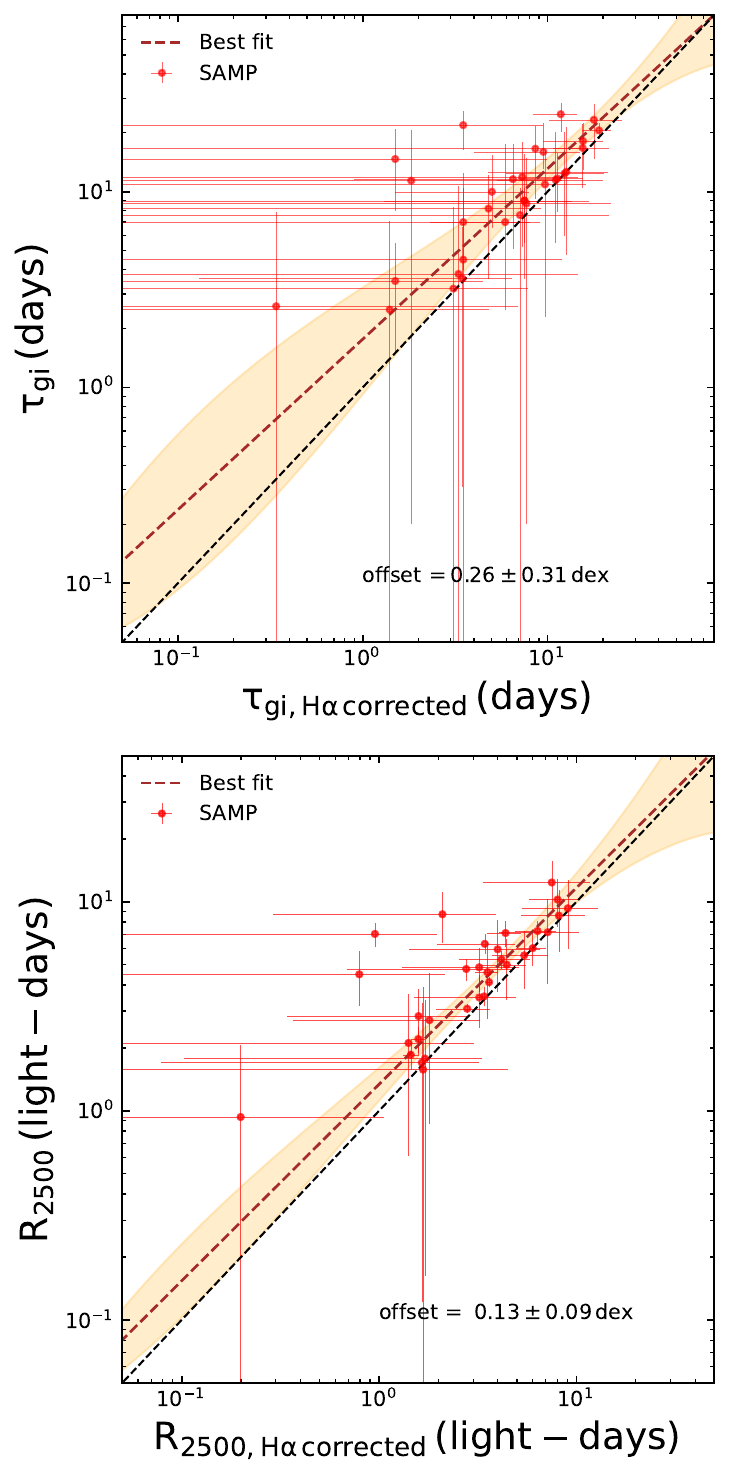}

\caption{ Top: Comparison of inter-band lags between the $g$ and $i$ bands ($\tau_{gi}$), without and with H$\alpha$ line contamination correction in the $i-$ band. Bottom: Comparison of disk sizes without and with H$\alpha$ contamination correction in the $i-$ band for SAMP targets. The dashed brown line represents the best-fit to the data points, while the dashed black line denotes the 1:1 relation in each panel.}
\label{fig:disk_halp}
\end{figure}

\begin{table*}[]
\centering
\movetableright= -12mm
 \caption{Linear regression fitting results of $R_{2500}$--$L_{5100}$, and $R_{\text{BLR}}$--$R_{2500}$ relations}
 \label{tab:fit}

\resizebox{18cm}{!}{
\begin{tabular}{lcccr} \hline \hline

Sample & relation & slope & intercept & $\sigma_{int}$ (dex)  
\\ 
(1) & (2) & (3) & (4) & (5) 
\\ \hline
SAMP + Literature & $R_{2500}$--$L_{5100}$ & $0.58\pm0.03$ & $0.55\pm0.04$ & 0.23 \\
SAMP + Literature (without NGC~4395) & '' & $0.55\pm0.05$ & $0.54\pm0.04$ & 0.23 \\
SAMP + Literature + Guo+22 & '' & $0.44\pm0.03$ & $0.38\pm0.03$ & 0.31 \\
SAMP + Literature + Guo+22 (without NGC~4395) & '' & $0.39\pm0.04$ & $0.38\pm0.02$ & 0.30 \\
\hline

SAMP + Literature & $R_{\text{BLR}}$--$R_{2500}$ & $0.87\pm0.07$ & $1.60\pm0.05$ & 0.29 \\
SAMP + Literature (without NGC~4395) & '' & $0.69\pm0.10$ & $1.59\pm0.05$ & 0.26 \\
SAMP + Literature + Guo+22 & '' & $0.87\pm0.07$ & $1.61\pm0.05$ & 0.30 \\
SAMP + Literature + Guo+22 (without NGC~4395) & '' & $0.69\pm0.10$ & $1.61\pm0.04$ & 0.28 \\

\hline

\end{tabular}

}
\parbox{\linewidth}{
        \vspace{1em} 
        \noindent
        \textbf{Note.} Columns are (1) sample used for fitting, (2) fitting relation, (3) slope obtained from best-fit, (4) intercept values from best-fit, and (5) intrinsic scatter.}

 
\end{table*}

\subsection{{$M_{\text{BH}}$ estimates using the BLR size -- $R_{2500}$ relation}}

 We explore the potential use of the BLR size -- $R_{2500}$ relation for estimating $M_{\text{BH}}$. Instead of using the direct measurement of the BLR size from H$\beta$ reverberation-mapping or utilizing the empirical BLR size -- luminosity relation \citep[e.g.,][]{2013ApJ...767..149B, 2024ApJ...962...67W, 2024ApJS..275...13W}, we can utilize the BLR size -- R$_{2500}$ relation (Equation \ref{eq_blr}) to infer the BLR size and estimate $M_{\text{BH}}$ after combining with the broad line velocity. 
Note that measuring H$\beta$ lags for high-luminosity AGNs at high-$z$ is challenging due to intrinsically large $R_{\text{BLR}}$ values and redshift effects. In the case of the H$\beta$ size -- luminosity relation, there is a systematic effect due to the Eddington ratio \citep[see discussion by][]{2024ApJ...962...67W}, not to mention that the BLR size -- luminosity relation is not very well established for C IV and Mg II lines for high-$z$ AGNs. 


We derive $R_{\text{BLR}}$ from $R_{2500}$ using Equation \ref{eq_blr}, while we obtain the velocity $\Delta V$ from H$\beta$ line dispersion ($\sigma_{\text{line}}$) for the SAMP sample \citep{2024ApJ...962...67W} and H$\beta$ (H$\alpha$ for NGC 4395) FWHM for the Literature sample from \citet{2019ApJ...886...42D} or the corresponding literature.
Then, we adopt a virial factor $f_{\text{BLR}}$ = 4.47 \citep[or 1.12 for FWHM;][]{2015ApJ...801...38W} to determine $M_{\text{BH, CRM}}$ = $f_{\text{BLR}} \big(\frac{R_{\text{BLR}} \Delta V^2}{G}\big)$.

We compare the continuum-RM based mass with the H$\beta$ reverberation based mass in Figure \ref{fig:SE_bh}, finding 
a sublinear relation with a slope of $\phi = 0.90$, which arises from the sublinear relation between the $R_{\text{BLR}}$ and $R_{2500}$ (Figure \ref{fig:blr_cer}). We find negligible offset of $0.02\pm0.06$ dex between $M_{\text{BH, CRM}}$ and $M_{\text{BH, RM}}$.

Nevertheless, we find an intrinsic scatter of $\sigma_{\text{int}} = 0.33$ dex between the two masses, which is comparable to the scatter reported by \citet{2020ApJ...903..112D}: 0.31 dex when using $\sigma_{\text{line}}$ and 0.37 dex when using FWHM. In contrast our measured scatter is slightly larger than the 0.23 dex scatter reported by \citet{2023ApJ...948L..23W}, who compared their estimated $M_{\text{BH, CRM}}$ with single-epoch $M_{\text{BH}}$ inferred from the H$\beta$ $R_{\text{BLR}}$ -- $L_{5100}$ relation. 
The empirical $R_{\text{BLR}}$ -- $R_{2500}$ relation provides the potential of estimating $M_{\text{BH}}$ once the systematic sublinear trend between $R_{\text{BLR}}$ and $R_{2500}$ is better investigated with a larger sample.  Future large-scale monitoring surveys, such as the Legacy Survey of Space and Time (LSST), will provide an unprecedented large sample of AGNs feasible for continuum-RM and combining with single-epoch spectroscopic follow-up, continuum-RM based black hole mass can be obtained, providing valuable opportunities for investigating black hole growth and accretion history.

\subsection{H$\alpha$ emission line contamination in ZTF$-i$ band}\label{sec:lgsp}
 
 The presence of the broad H$\alpha$ emission line in the ZTF bands can affect continuum lag measurements, as well as the disk size estimates from the lag-spectra. To test the effect of this excess, we perform a test by removing the contribution of H$\alpha$ based on the following steps. First, we search for the SDSS spectrum of each AGN in our sample to measure the H$\alpha$ flux and its contribution in the $i$-band. We obtain spectra for 33 out of 37 targets. We decompose each spectrum with a power-law continuum, the {\FeII} template from \citet{1992ApJS...80..109B}, and an H$\alpha$ line profile with three Gaussians for a broad component and one Gaussian for a narrow component. Using the modeled H$\alpha$ line (including narrow components) and the ZTF$-i$ filter transmission curve in the observed-frame, we calculate the fraction of H$\alpha$ flux in the $i-$ band. 

Second, to construct an H$\alpha$-line-free $i-$band continuum light curve, we assume the H$\alpha$ time lag ($\tau_{\text{H}\alpha}$) using the $R_{\text{H}\alpha} - L_{5100}$ relation from \citet{2023ApJ...953..142C}. We then generate an H$\alpha$ light curve by multiplying the H$\alpha$ fraction to the observed $i-$ band light curve, and shifting the resulting light curve by the difference between the assumed H$\alpha$ lag and the inter-band lag between $g$ and $i-$ bands (i.e., $\tau_{\text{H}\alpha} - \tau_{gi}$,). 
After that we apply DRW modeling to construct the mock H$\alpha$ light curve. Finally, we subtract the mock H$\alpha$ line light curve from the observed $i-$ band light curve to obtain the H$\alpha$-line-free continuum $i-$ band light curve. Note that there are large seasonal gaps in the ZTF light curves, which leads to large uncertainties in the simulated H$\alpha$ light curve. 
Finally, we derive lags between the $g$ and H$\alpha$-subtracted continuum $i-$ band light curves, using ICCF analysis followed by lag-spectrum fitting. 

We present the distribution of the fraction of H$\alpha$ flux in the ZTF$-i$ band in Figure \ref{fig:halp_dist} (left). The H$\alpha$ contamination in the $i-$ band ranges from $\sim$ 0.0 \% to 42.5 \%, with a median value of 28.8 \%. We expect that targets with a larger H$\alpha$ fraction could have a larger correction of $\tau_{gi}$. Figure \ref{fig:halp_dist} (right) shows the estimated correction fraction of $\tau_{gi}$ as a function of the H$\alpha$ fraction in the $i-$ band, indicating that the expected trend is not clear, presumably due to the large uncertainty of the H$\alpha$ correction. Note that the simulated H$\alpha$ flux is not reliable, if it is generated in the seasonal gap of the $i-$ band light curve. 

In Figure \ref{fig:disk_halp} we compare the $\tau_{gi}$ values (top) and the disk size (bottom) with and without the H$\alpha$ correction. While a few targets show a large correction, the unertainty of the lag measurements are substantially large. We find that $\tau_{gi}$ is overestimated by an average factor of 1.8 ($0.26 \pm 0.31$ dex) and the disk size by a factor of 1.3 ($0.13 \pm 0.09$ dex) without correcting for the contribution of the broad H$\alpha$ emission in the observed ZTF$-i$ band. However, the slopes in the $R_{2500}$--$L_{5100}$ relation ($\alpha = 0.54 \pm 0.04$) and the $R_{\text{BLR}}$--$R_{2500}$ relation ($\gamma = 0.89 \pm 0.08$) remain rather consistent with our previous values within uncertainties, even after accounting for H$\alpha$ contamination. Note that for the given large uncertainties of the simulated H$\alpha$ light curves, we present this result as a test. Further detailed studies are required to clearly investigate the effect of H$\alpha$ contribution on the slope of the size--luminosity relation.

\section{Summary}
\label{sec:summ}

 We present continuum reverberation-mapping results for 37 high-luminosity AGNs selected from the SAMP, utilizing light curves in the $B$, $V$, and ZTF $g$,$r$,$i$ bands. 
 A thorough quality assessment ensures the consistency and reliability of inter-band lag measurements among the selected targets. 
By including the literature sample, we expand the dynamical range of AGN luminosity, BH mass, and Eddington ratios. Our main findings are summarized below. 

\begin{itemize}
    \item While the inter-band lags increase with wavelength, the disk size estimated from the lag spectrum fitting exceeds the prediction from the SSD model (assuming $\chi = 2.49$) by approximately 4 to 6 times (i.e., $0.69 \pm 0.04$ dex). These findings align with the results of previous studies with disk RM and microlensing.


    \item The continuum emitting region size at 2500 {\AA} correlates with the AGN monochromatic luminosity at 5100\AA, $L_{5100}$ as $R_{2500} \, \propto \, L_{5100}^{\alpha}$, with a slope of $\alpha = 0.58 \pm 0.03$. This suggests that $R_{2500}$ is primarily driven by $L_{5100}$, with evidence pointing to a significant contribution from BLR diffuse continuum emission to the observed continua. However, the observed slope is steeper than the value of 0.5  expected from the photo-ionization of BLR gas clouds. When combining this observed relationship with the luminosity--mass dependency within our sample, we find that $R_{2500}$ consistently depends on $M_{\text{BH}} \times L_{5100}$, as observed and as predicted by the  SSD model. This may indicate a combined contribution of both disk emission and BLR diffuse continuum emission while the disk emission in dominant.


    \item A strong correlation between the BLR size and continuum emitting region size is found as $R_{\text{BLR}} \, \propto \, R_{2500}^{\gamma}$ with a slope of $\gamma=0.87\pm0.07$, indicating that the BLR size is $\sim1.25\pm0.12$ dex larger than that of $R_{2500}$, a finding consistent with observations  and the model predicted value. Establishing this relationship across a wide dynamic range in luminosity holds particular significance for estimating single-epoch BH masses in a large sample of AGNs, especially in light of future astronomical surveys, such as LSST.

\end{itemize}


\begin{acknowledgments}

 We thank the anonymous referee for valuable comments and suggestions. This work has been supported by the Basic Science Research Program through the National Research Foundation of the Korean Government (grant No. NRF-2021R1A2C3008486 and 2019R1A6A1A10073437). A.K.M. acknowledges the support from the European Research Council (ERC) under the European Union's Horizon 2020 research and innovation program (grant No. 951549). 

This paper uses data from the Zwicky Transient Facility (ZTF). ZTF is funded by the National Science Foundation through grant number AST-1440341 and a partnership with Caltech, IPAC, the Weizmann Institute for Science, the Oskar Klein Center at Stockholm University, the University of Maryland, the University of Washington, Deutsches Elektronen-Synchrotron and Humboldt University, Los Alamos National Laboratories, the TANGO Consortium of Taiwan, the University of Wisconsin at Milwaukee, and Lawrence Berkeley National Laboratories. COO, IPAC, and UW manage operations.

    {\bf Software:}  {\tt JAVELIN} \citep{2011ApJ...735...80Z}, {\tt PyCALI} \citep{2014ApJ...786L...6L}\\
\end{acknowledgments}

%

\vspace{0.01mm}
\facilities{IRSA}





\appendix \label{app}

\section{Consistency check of lag analysis between ICCF and {\tt JAVELIN}}
\label{sec:sim}

 To evaluate the consistency of lag analysis between ICCF and {\tt JAVELIN}, we perform a simulation. We generate the driving irradiating light curve, C(t), by incorporating both large-scale and small-scale flux variations, representing intrinsic AGN variability. Large-scale variations are modeled as random fluctuations, while small-scale variations follow a cumulative random process, reflecting the stochastic nature of AGN continuum emission. The light curve is simulated with a 2-day cadence, an SNR of 100, and a 3-year baseline, consistent with the real data used in our analysis. A typical seasonal gap of 90 days is introduced in C(t) to mimic observational constraints.

\restartappendixnumbering

\begin{figure*}
\centering
\hspace{-1.2cm}
\includegraphics[scale=0.6]{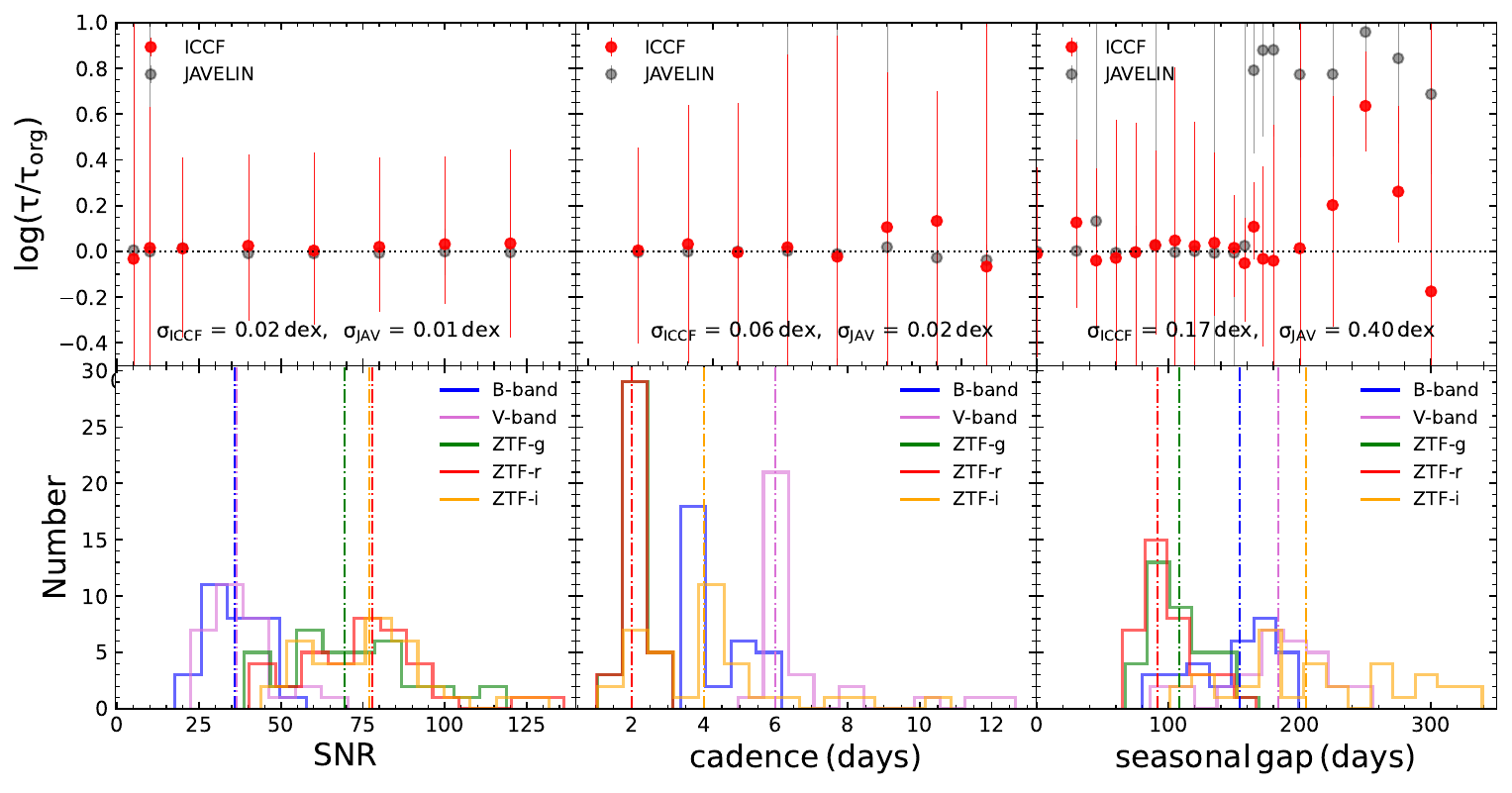}

\caption{Top: Lag ratio of the recovered to the original time delay obtained using ICCF (red) and {\tt JAVELIN} (black) as a function of SNR (left), cadence (middle), and seasonal gap (right), with standard deviations indicated. The horizontal black dotted line shows 1:1 relation.  Bottom:  Distributions of SNR, median cadence, and seasonal gaps for light curves in the $B$, $V$, $g$, $r$, and $i-$ bands, with median values marked by vertical dash-dotted lines in the corresponding panels.}
\label{fig:sim_icfjav}
\end{figure*}

The responding light curve, I(t), is derived from C(t) with a time lag of $\tau_{org} = 5$ days. This is achieved by interpolating and scaling the flux in C(t) at lagged time points while introducing irregular cadence variations between 2 and 4 days, reflecting real survey conditions. Additionally, a top-hat smoothing kernel is applied to account for the extended reprocessing response.

We then recover the time delays for different SNRs, cadences, and seasonal gaps using ICCF and {\tt JAVELIN}. Notably, cadence variations are introduced only in I(t) when performing lag recovery at different cadences, while SNR and seasonal gap changes in both light curves preserve the original cadence values. The recovered time delays ($\tau$) are compared to the original $\tau_{org}$ as a function of these parameters, as shown in the top panel of Figure \ref{fig:sim_icfjav}. The bottom panel displays the  distributions of SNR, cadence, and seasonal gaps for the observed light curves in the $B$, $V$, $g$, $r$, and $i$ bands used in our analysis.

Our simulation demonstrates that within the observed range of SNR, both ICCF and {\tt JAVELIN} successfully recover inter-band time delays. However, when cadence exceeds twice the expected lag, the recovered lags exhibit minor deviations, though they remain consistent within larger errors. This reliability is maintained because the driving light curve, C(t), remains densely sampled and unchanged throughout the simulation. Regarding seasonal gaps, ICCF and {\tt JAVELIN} recover the original lag for seasonal gaps shorter than 200 and 150 days, respectively. Beyond these thresholds, lag estimates become increasingly uncertain, with {\tt JAVELIN} showing more pronounced deviations, exhibiting a scatter of 0.40 dex compared to 0.17 dex for ICCF.

Although both methods yield consistent lags in most cases, {\tt JAVELIN} fails to accurately recover time delays for light curves with large seasonal gaps ($> 150$ days). A similar inconsistency in {\tt JAVELIN} compared to ICCF and {\tt PyROA} \citep{2021MNRAS.508.5449D} was recently reported by \citet{2024ApJS..275...13W}. Given these findings, we adopt ICCF lags for our analysis.

\section{Lag-spectrum for the SAMP and Literature AGNs} \label{app:lag_spc}

\begin{figure*}[!h]
\centering
\includegraphics[scale=0.23]{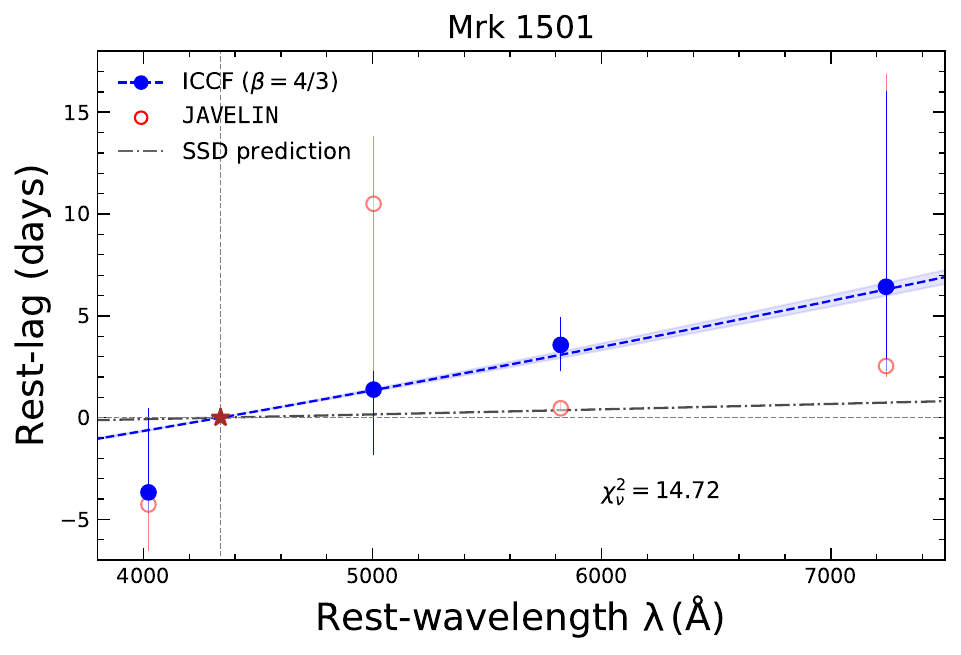}
\includegraphics[scale=0.23]{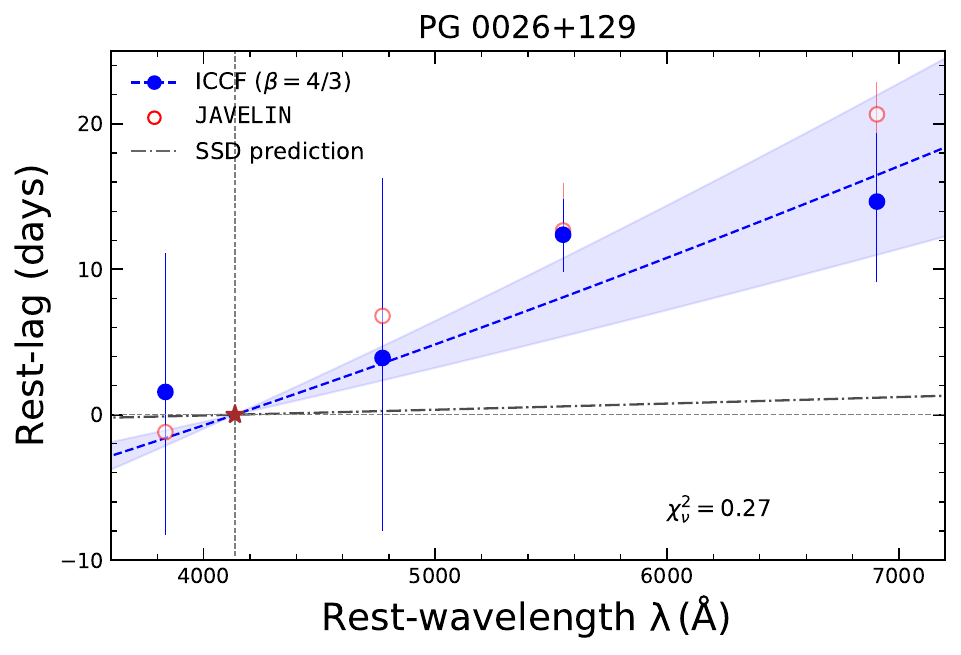}
\includegraphics[scale=0.23]{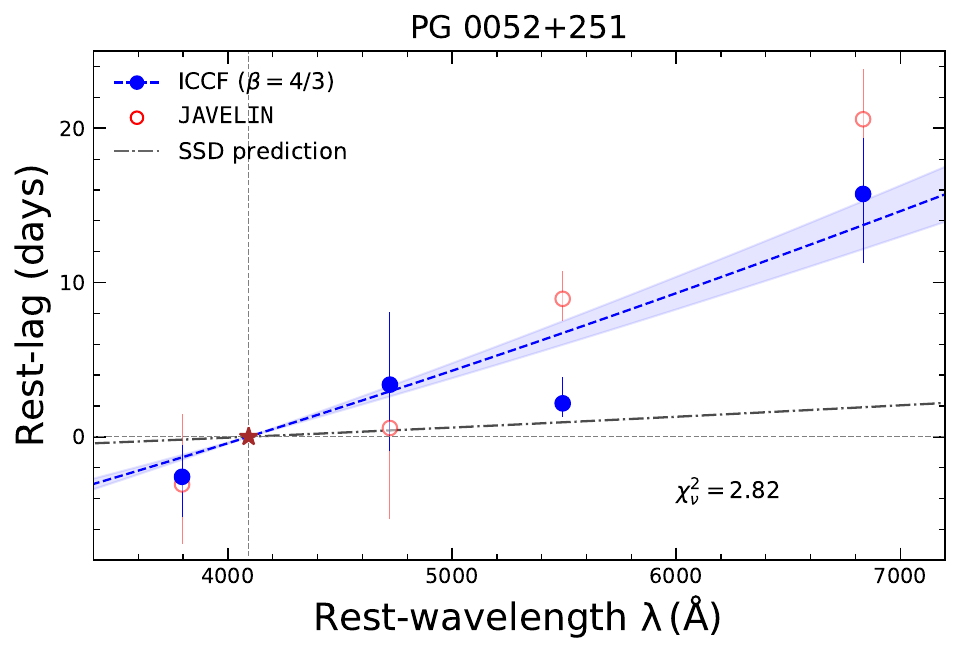}
\includegraphics[scale=0.23]{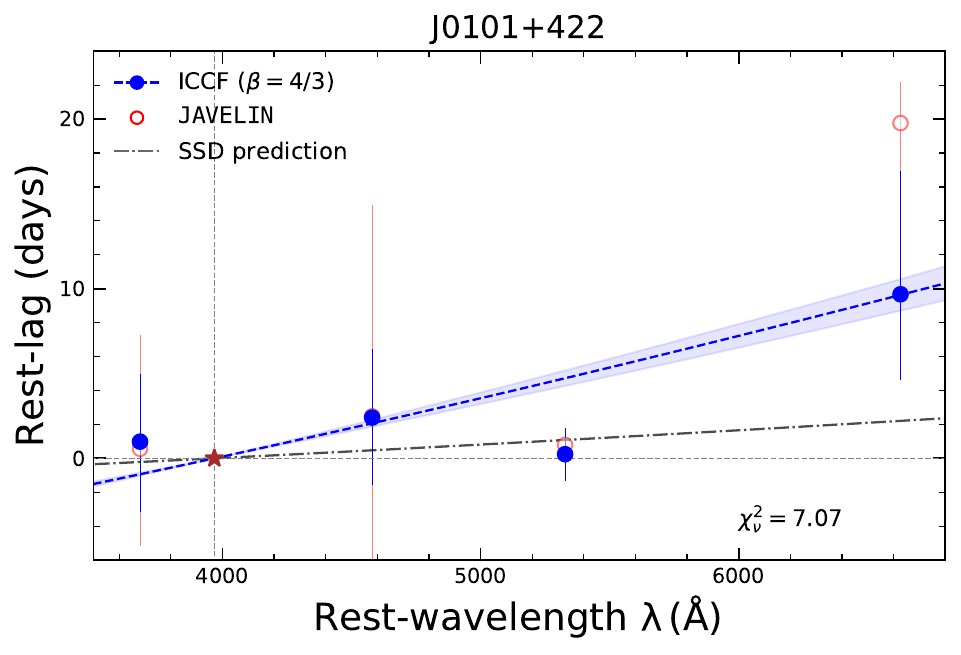}
\includegraphics[scale=0.23]{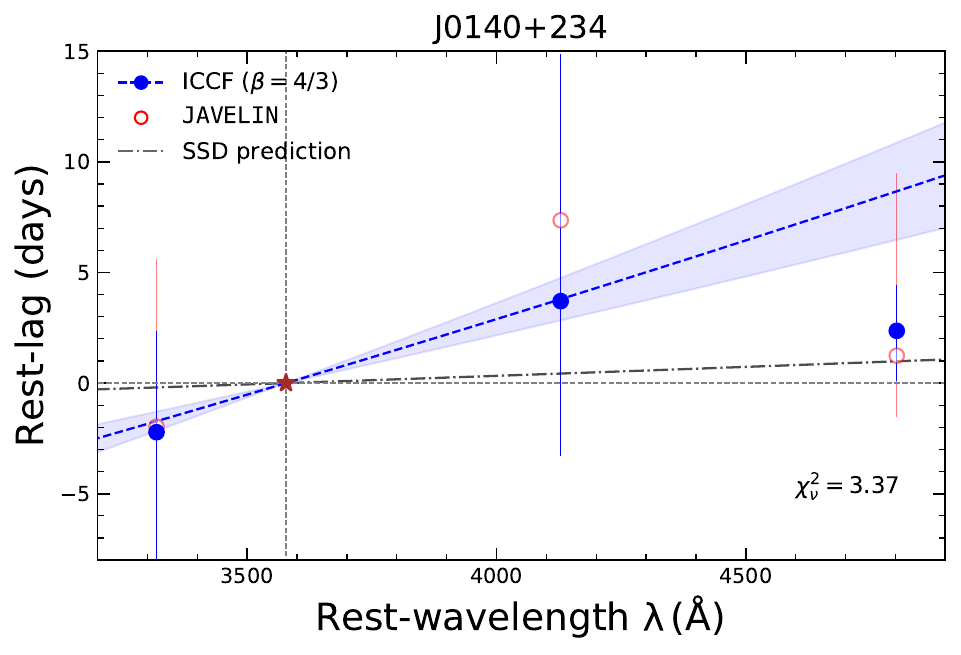}
\includegraphics[scale=0.23]{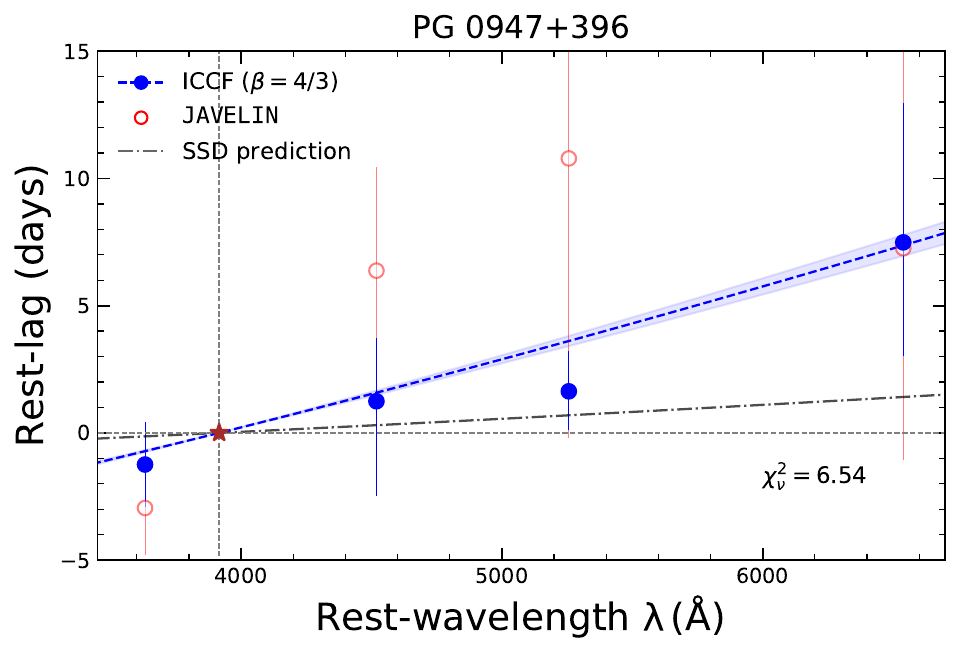}
\includegraphics[scale=0.23]{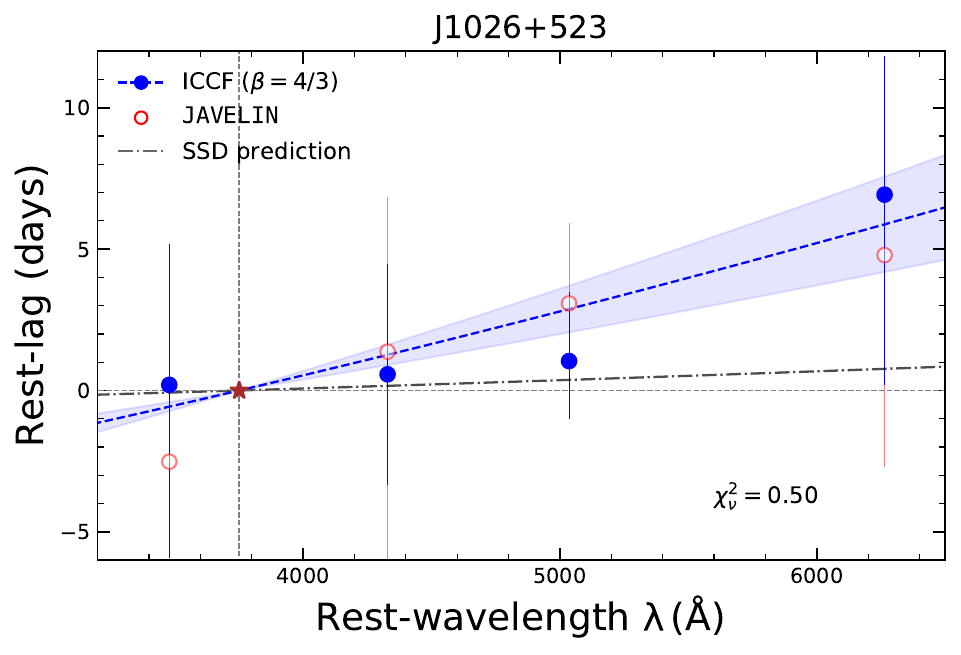}
\includegraphics[scale=0.23]{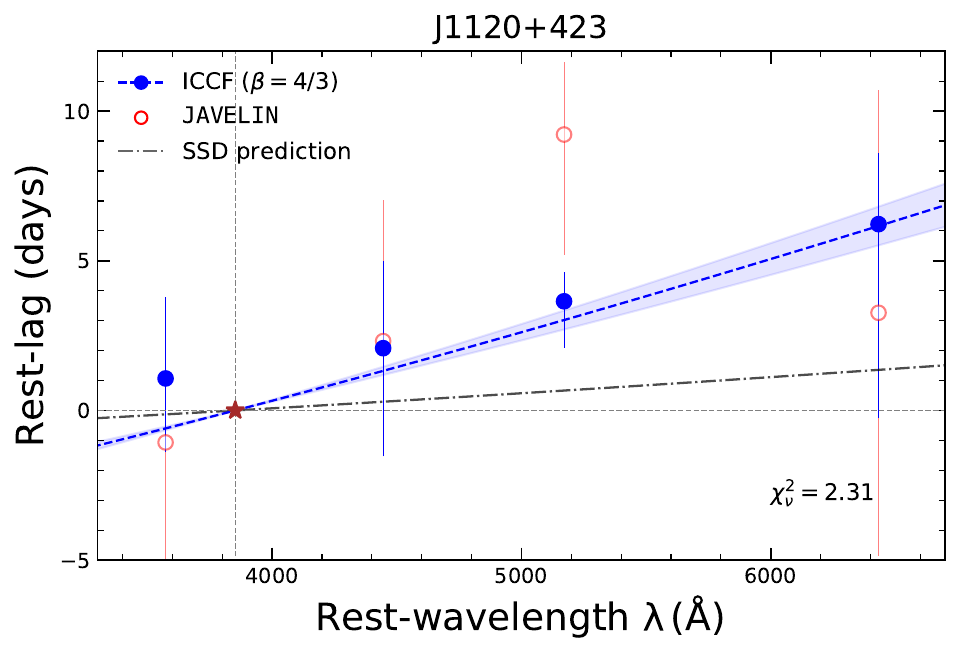}
\includegraphics[scale=0.23]{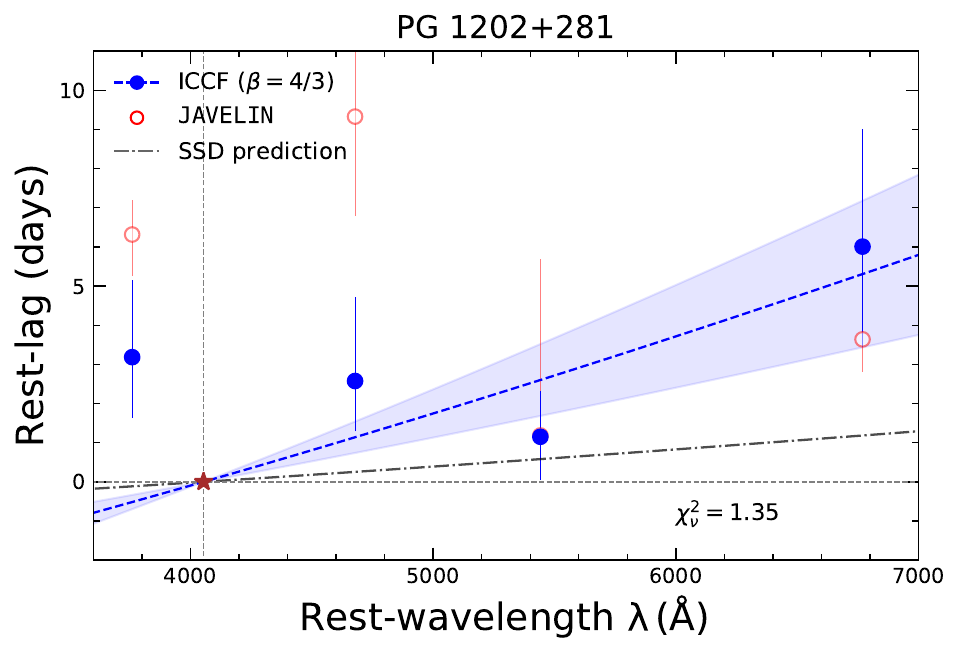}
\includegraphics[scale=0.23]{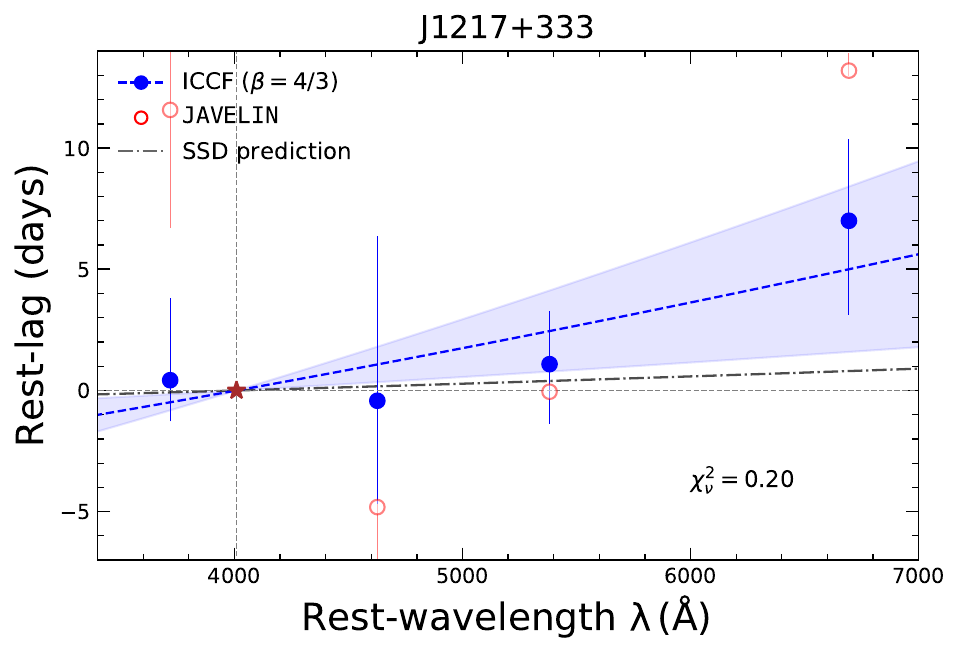}
\includegraphics[scale=0.23]{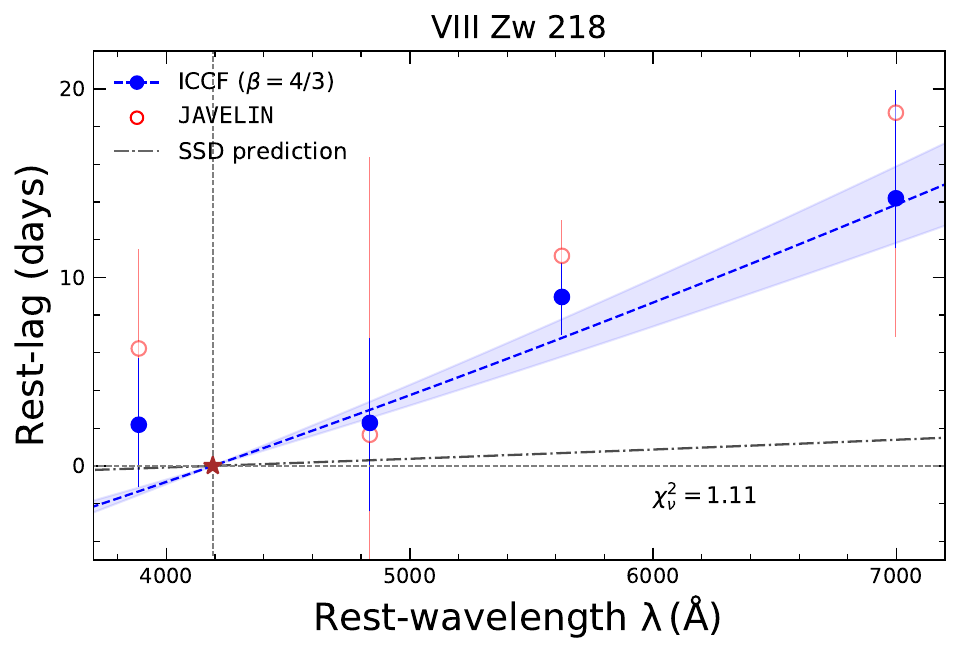}
\includegraphics[scale=0.23]{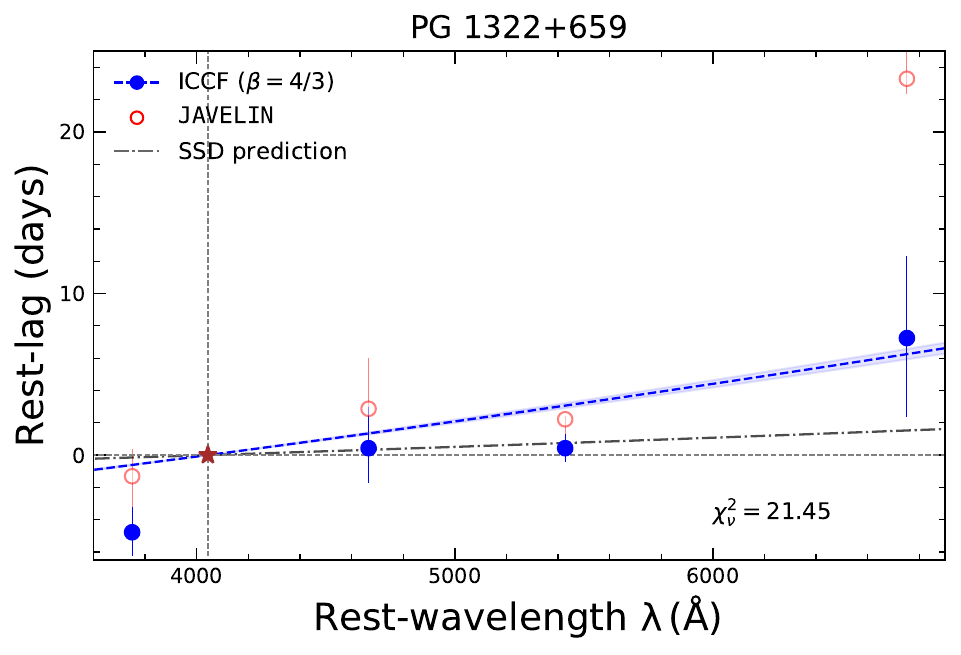}

\caption{Lag-spectrum: rest-frame lags as a function of rest-frame wavelength for SAMP sample. Same notations are used as Figure \ref{fig:lagspec}. The full set of lag-spectrum plots for the SAMP targets is available online.
}
\label{fig:lagspc1}
\end{figure*}

\begin{figure*}[!h]
\centering
\includegraphics[scale=0.23]{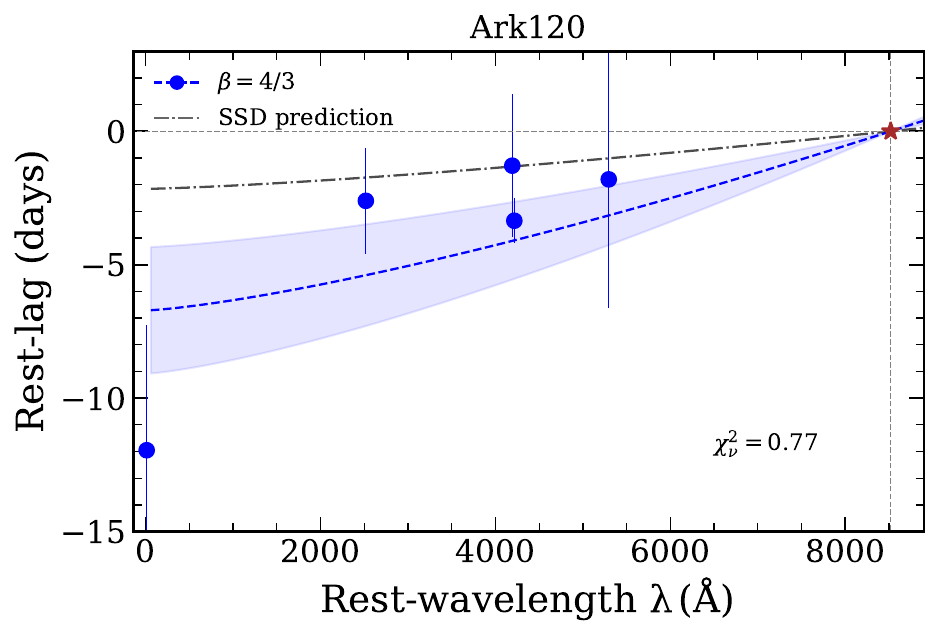}
\includegraphics[scale=0.23]{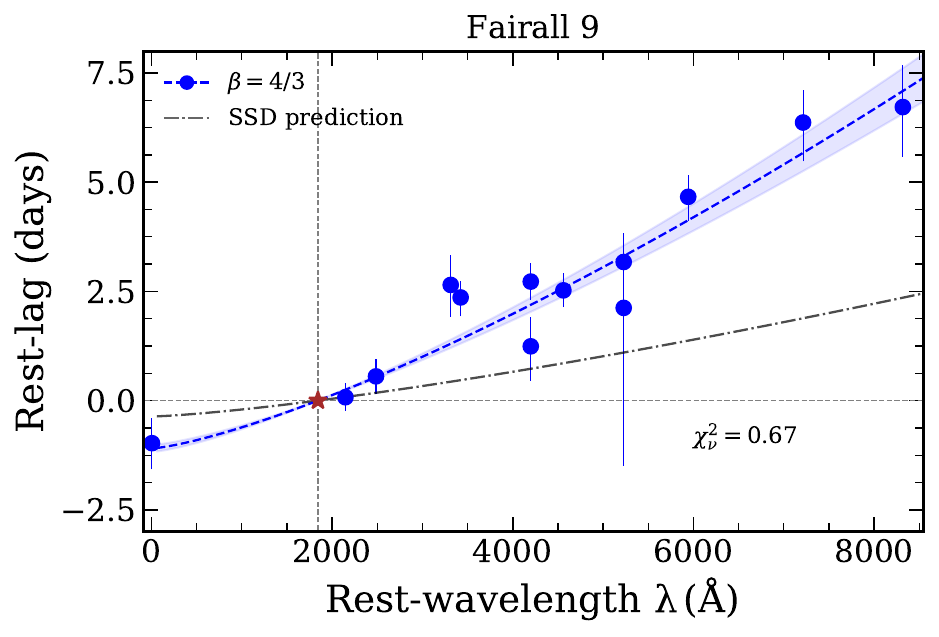}
\includegraphics[scale=0.23]{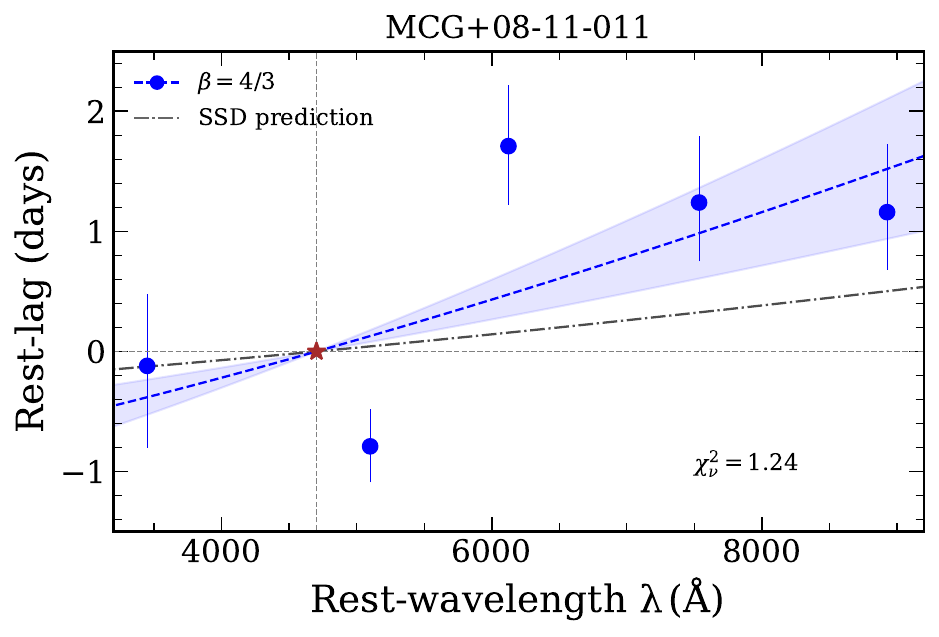}
\includegraphics[scale=0.23]{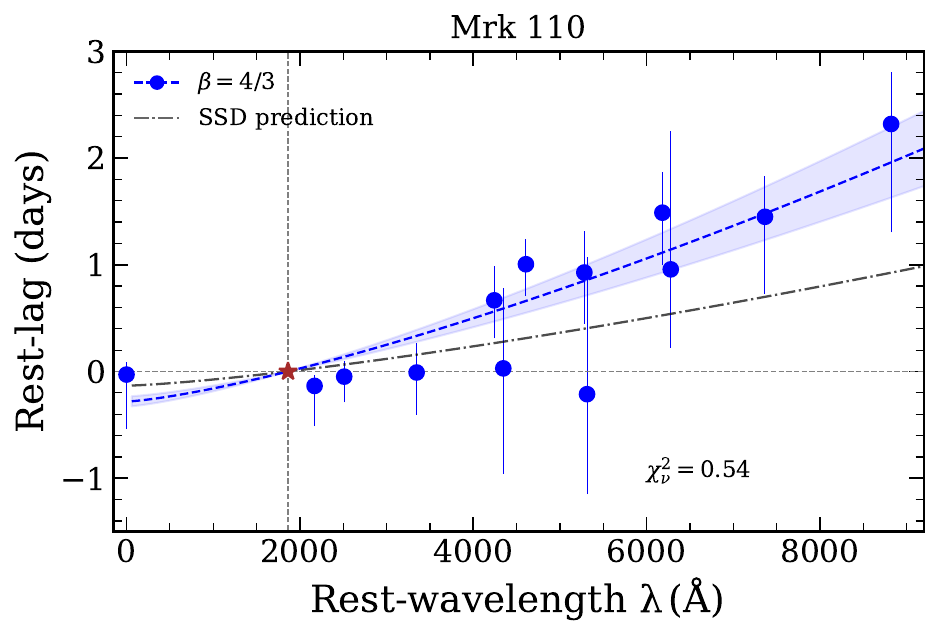}
\includegraphics[scale=0.23]{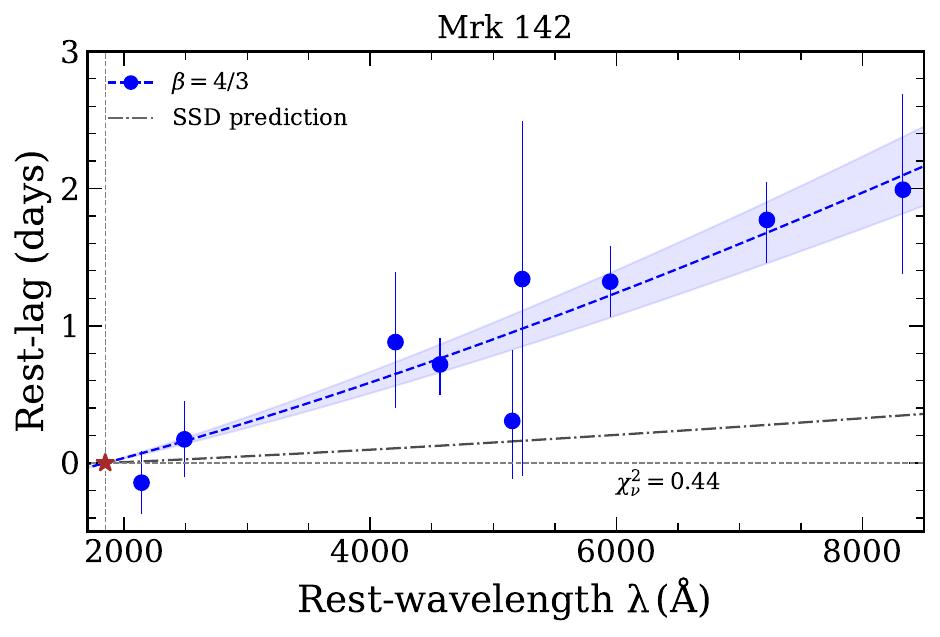}
\includegraphics[scale=0.23]{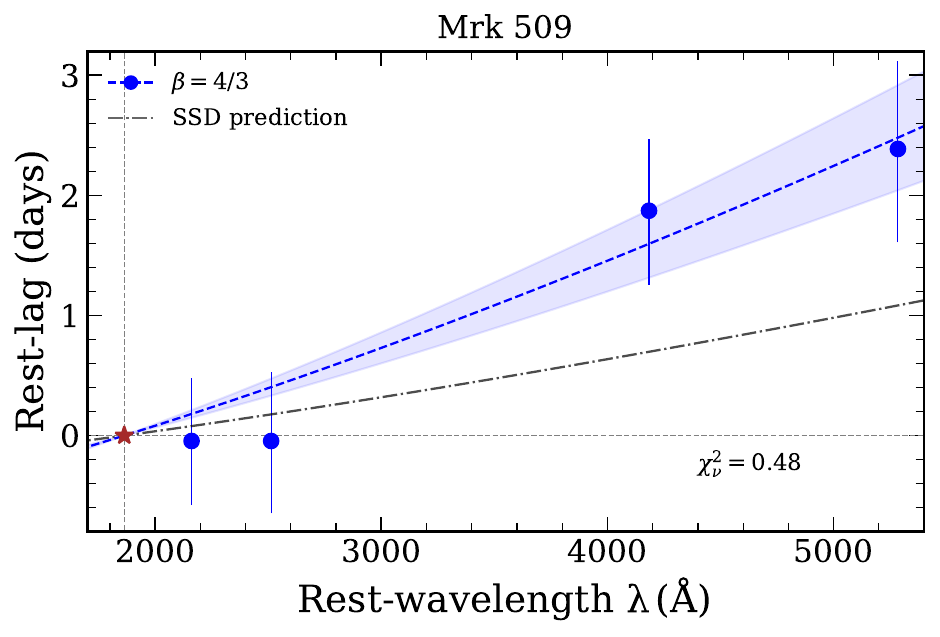}
\includegraphics[scale=0.23]{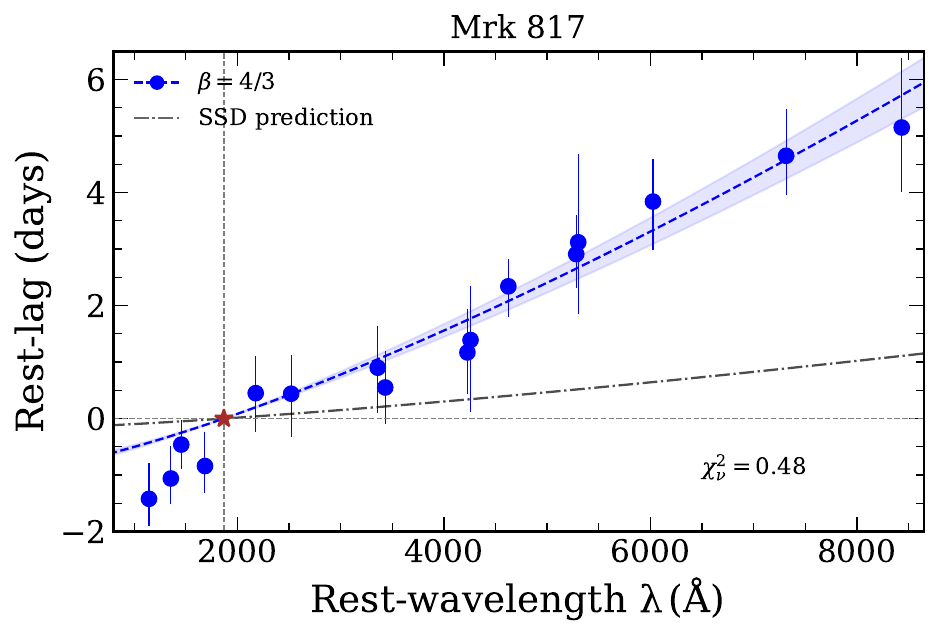}
\includegraphics[scale=0.23]{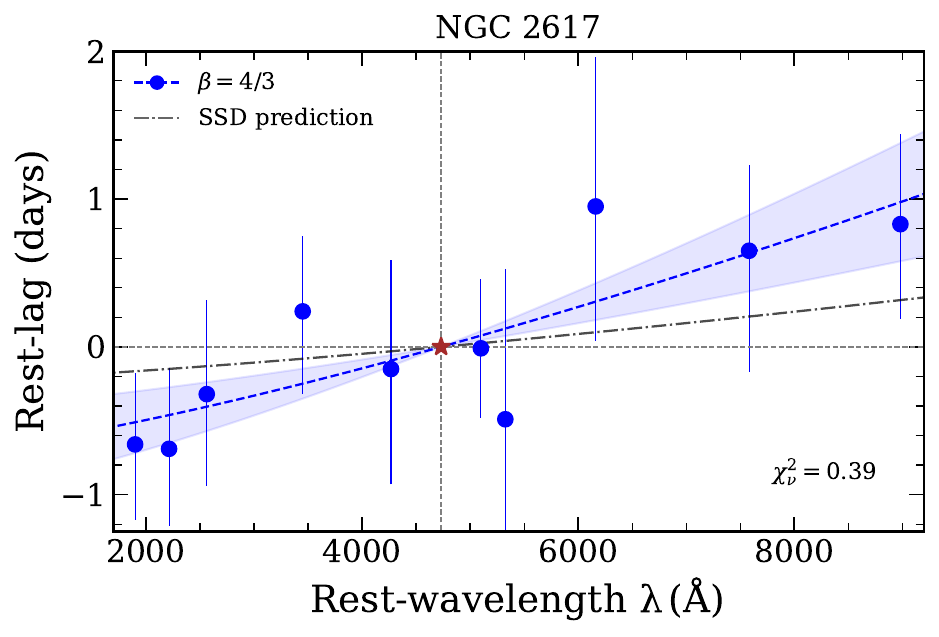}
\includegraphics[scale=0.23]{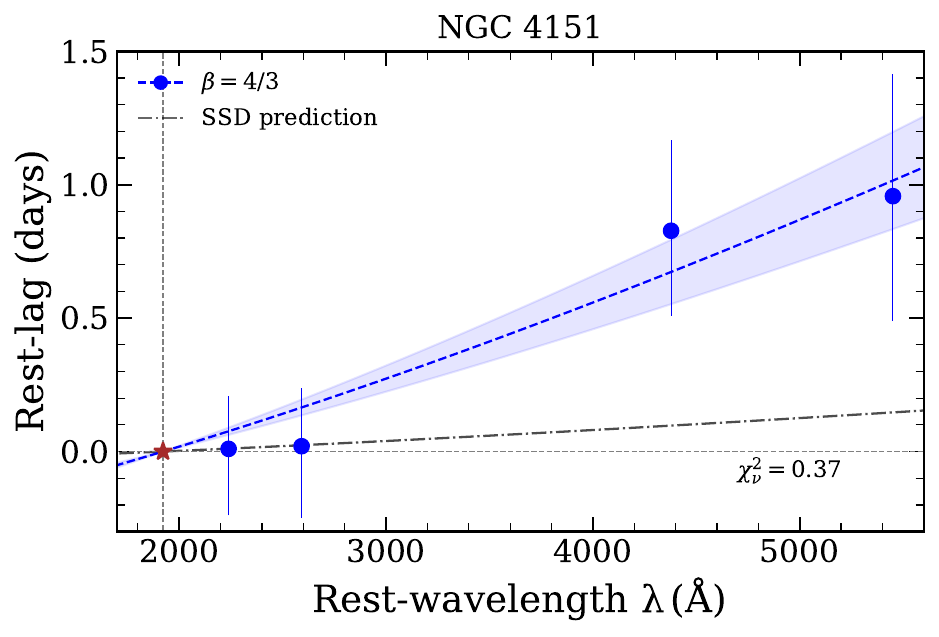}
\includegraphics[scale=0.23]{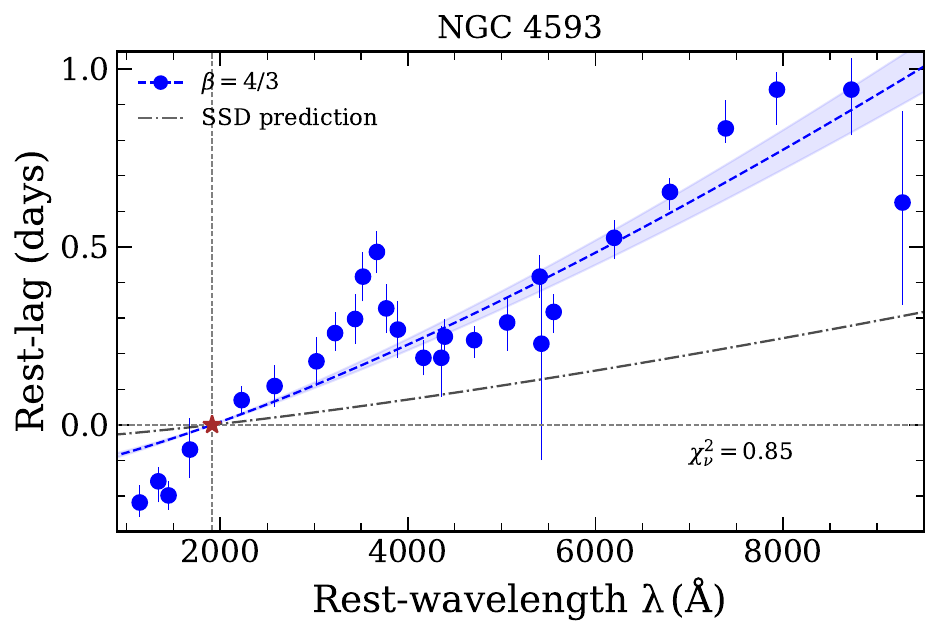}
\includegraphics[scale=0.23]{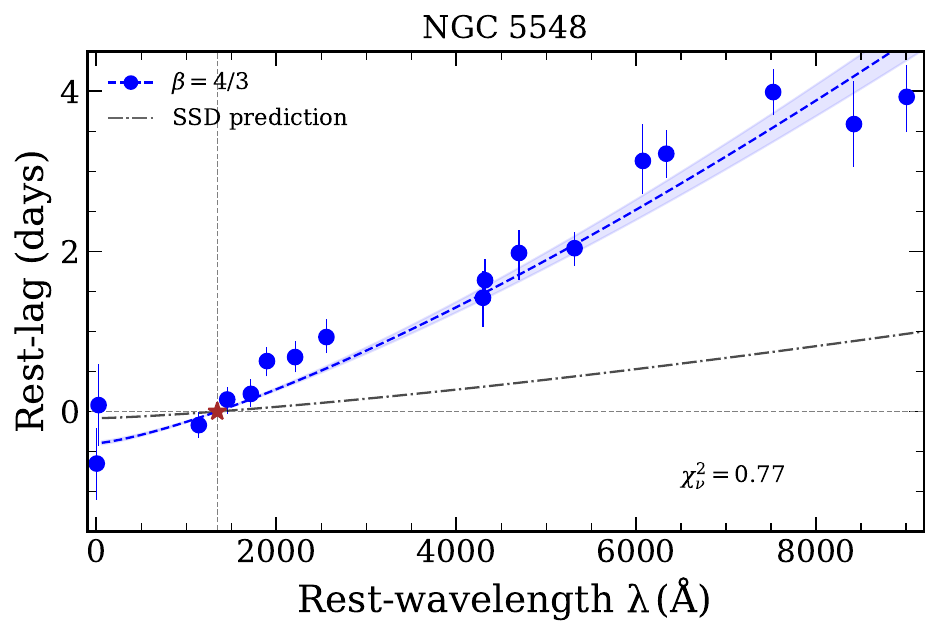}
\includegraphics[scale=0.23]{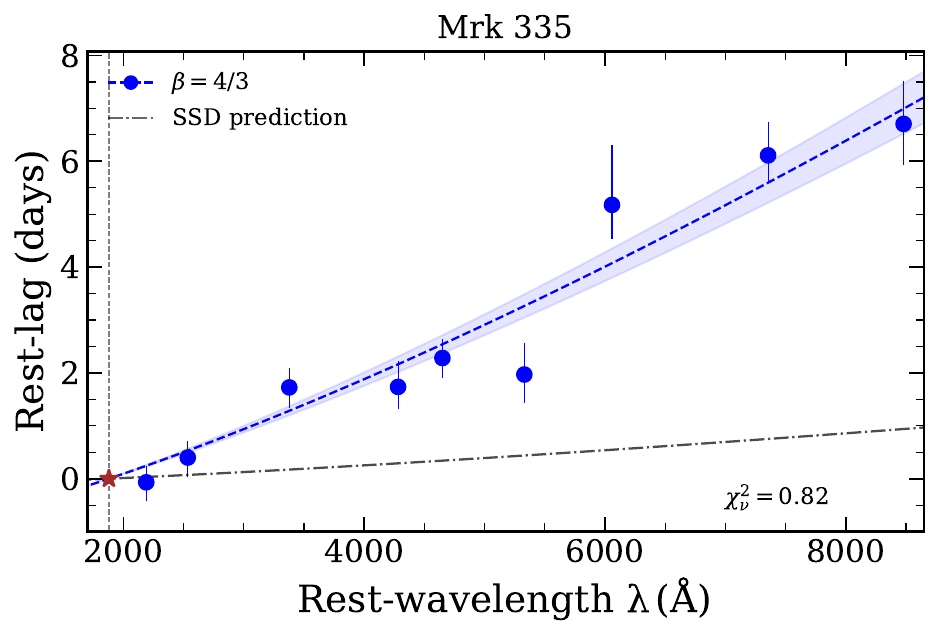}

\caption{Lag-spectrum: rest-frame lags as a function of rest-frame wavelength for Literature sample (blue circular points). The dashed blue line represents the best-fitting $\tau \, \propto \, \lambda^{4/3}$ to the lags with shaded region indicating the $1\sigma$ uncertainty from the fitting. The dashed-dotted black line represents the lag-wavelength dependence expected from SSD. The brown star indicates the reference point. The vertical and horizontal dotted lines denote the rest-frame reference wavelength and zero rest-lag, respectively. Additionally, the $\chi^2$ per degree of freedom ($\chi^2_{\nu}$) is presented. The full set of lag-spectrum plots for the Literature targets is available online.}
\label{fig:lagspc_lit}
\end{figure*}

Continuum lags measured with respect to ZTF$-g$ band in the rest-frame are shown as a function of rest-frame wavelength for the remaining 36 SAMP targets in Figure \ref{fig:lagspc1}. The rest-frame lags are fitted using Equation \ref{fit_eq} with $\beta=4/3$ to derive CER size at the rest-frame reference wavelength, i.e., ZTF$-g$ band at 4722.74{\AA}/(1 + $z$). The corresponding lag-spectra for the targets from Literature are shown in Figure \ref{fig:lagspc_lit}, where the reference point in the rest-frame is shown by the brown star. All lag-spectra are obtained in the rest-frame.

\restartappendixnumbering

\bibliography{sample631}{}
\bibliographystyle{aasjournal}



\end{document}